\documentclass[twoside,twocolumn]{article}

\usepackage{graphicx}

\usepackage[sc]{mathpazo} 
\usepackage[T1]{fontenc} 
\fontfamily{qcr}\selectfont
\usepackage{microtype} 

\usepackage[english]{babel} 

\usepackage[hmarginratio=1:1,top=32mm,columnsep=20pt]{geometry} 
\usepackage[hang, small,labelfont=bf,up,textfont=it,up]{caption} 
\usepackage{booktabs} 

\usepackage{abstract} 

\usepackage{titlesec} 
\renewcommand\thesection{\Roman{section}} 
\renewcommand\thesubsection{\roman{subsection}} 
\titleformat{\section}[block]{\large\scshape\centering}{\thesection.}{1em}{} 
\titleformat{\subsection}[block]{\large}{\thesubsection.}{1em}{} 

\usepackage{fancyhdr} 
\usepackage{titling} 

\usepackage{hyperref} 

\pretitle{\begin{center}\Huge\bfseries} 
\posttitle{\end{center}} 
\title{Molecular Simulations for the Spectroscopic Detection of Atmospheric Gases} 
\author{%
\textsc{Clara Sousa-Silva}\thanks{Corresponding author; {\bf{cssilva@mit.edu}}.} \thanks{Electronic Supplementary Information (ESI) available and routinely updated on GitHub/csousasilva/RASCALL and \cite{rascall_database}.} \\[1ex] 
\normalsize Massachusetts Institute of Technology \\ 
\normalsize Department of Earth, Atmospheric, and Planetary Sciences  \\ 
\normalsize Department of Physics \\ 
\and 
\textsc{Janusz J Petkowski} \\[1ex] 
\normalsize Massachusetts Institute of Technology \\ 
\normalsize Department of Earth, Atmospheric, and Planetary Sciences  \\ 
\and 
\textsc{Sara Seager} \\[1ex] 
\normalsize Massachusetts Institute of Technology \\ 
\normalsize Department of Earth, Atmospheric, and Planetary Sciences  \\ 
\normalsize Department of Physics  \\ 
\normalsize Department of Aeronautics and Astronautics \\
}
\date{\today} 
\renewcommand{\maketitlehookd}{%
\begin{abstract}

Unambiguously identifying molecules in spectra is of fundamental importance for a variety of scientific and industrial uses. Interpreting atmospheric spectra for the remote detection of volatile compounds requires information about the spectrum of each relevant molecule. However, spectral data currently exist for a few hundred molecules and only a fraction of those have complete spectra (e.g. H$_2$O, NH$_3$). Consequently, molecular detections in atmospheric spectra remain vulnerable to false positives, false negatives, and missassignments. There is a key need for spectral data for a broad range of molecules. Given how challenging it is to obtain high-resolution molecular spectra, there is great value in creating intermediate approximate spectra that can provide a starting point for the analysis of atmospheric spectra.

Using a combination of experimental measurements, organic chemistry, and quantum mechanics, RASCALL (Rapid Approximate Spectral Calculations for ALL) is a computational approach that provides approximate spectral data for any given molecule, including thousands of potential atmospheric gases. RASCALL is a new theoretical chemistry method for the simulation of spectral data.

RASCALL 1.0, presented here, is capable of simulating   molecular spectral data, in a few seconds, by interpreting functional group data from experimental and theoretical sources to estimate the position and strength of molecular bands. The RASCALL 1.0 spectra consist of approximate band centers and qualitative intensities.
 RASCALL 1.0 is also able to assess hundreds of molecules simultaneously, which will inform prioritization protocols for future, computationally and experimentally costly, high-accuracy physical chemistry studies. Finally, RASCALL can be used to study spectral patterns between molecules, highlighting ambiguities in molecular detections and also directing observations towards spectral regions that reduce the degeneracy in molecular identification.

 The RASCALL catalogue, and its preliminary version RASCALL 1.0, contains spectral data for more molecules than any other publicly available database, with applications in all fields interested in the detection of molecules in the gas phase (e.g., medical imaging, petroleum industry, pollution monitoring, astrochemistry). The preliminary catalogue of molecular data and associated documentation are freely available online and will be routinely updated. 
\end{abstract}
}

\begin{document}

\maketitle


\section{Introduction}
Spectral interpretation is indispensable to chemistry. Infrared (IR) spectroscopy, in particular, is one of the oldest and most well studied areas of chemical research. Since its discovery in 1800 \cite{herschel}, hundreds of compounds have been measured through IR spectroscopy, but it was not until 1937 that the first operating spectrometer was built \cite{firstspec}, heralding a golden age of IR spectroscopy. That golden age has now ended, as most molecular interpretation can be done with mass spectrometry or nuclear magnetic resonance spectroscopy. Nonetheless, IR spectroscopy remains a valuable tool for spectral analyses, in particular for the remote detection of volatile compounds.

 Infrared spectroscopy can be used in a wide-range of industrial applications such as climate monitoring\cite{ipcc, green1998imaging},  combustion analysis\cite{volin2001midwave, gorman2010generalization}, fire detections\cite{roberts2003evaluation}, biomedical imaging\cite{alabboud2007new}, pharmacology \cite{kupcewicz2013evaluation}, pollution tracking\cite{remer2008global}, atmospheric surveys \cite{ace}, geochemical mapping\cite{baugh1998quantitative}, catalytic cracking \cite{lopez2018monitoring}, and crop maintenance\cite{dong2013analyzing}. Spectral interpretation is also indispensable to astrochemistry, such as for studies of galaxies \cite{moustakas2010optical,spoon2006mid,evans2006dense}, stars\cite{kwan1982}, comets\cite{brooke1996}, brown dwarfs\cite{burrows1999chemical,allard2012models}, Solar System bodies\cite{bezard1990deep,hanel1972investigation,krasnopolsky2001detection,encrenaz1999atmospheric} and exoplanet atmospheres\cite{tinetti2007water,madhusudhan2009temperature, barman2015simultaneous,tsiaras2018population}.

Availability of fundamental spectral data is vital for the interpretation of observational spectra and the remote detection of individual molecules. There are thousands of molecular candidates that can contribute towards an atmospheric spectrum. Ideally, the spectrum of each of these molecules is completely known and freely available. For a few molecules (e.g. H$_2$O, NH$_3$, PH$_3$, CH$_4$ and CO$_2$), the available spectral data are of high quality: accurate, comprehensive and complete. However, thousands of molecules of interest have spectra that are still incorrect, incomplete, or completely unknown.
 
Spectra, or line lists, can be obtained by either experimental measurements or theoretical simulations; often a combination of both. Measuring molecular spectra experimentally leads to accurate spectra but it can be expensive, dangerous if the chemical is toxic, and sometimes altogether impossible. Instrumental limitations lead to whole regions of spectra being absent from measurements, and intensity values are often unreliable and difficult to extrapolate to non-measured temperatures. Theoretical calculations have become a valuable tool in obtaining molecular spectra but, for the majority of molecules, this process remains either computationally prohibitive or extremely slow. For example, the quantum chemical calculation of the complete molecular line list for methane \cite{yurchenko2014exomol}, which contains 35 billion transition features, took several years and required 4.5 million CPU hours, with the diagonalization of individual matrices taking up to 25 CPU hours, hundreds of cores and dozens of Tb of disk space. Spectra for molecules with more than 4 atoms are exponentially more difficult to simulate completely. The value of high accuracy quantum chemical simulations is undeniable but, as a consequence of their difficulty, spectra for thousands of molecules remain absent.

As new research fields emerge and discoveries become more ambitious, more molecular spectra will be needed. Traditional methods for obtaining line lists cannot be expected to address all of the thousands of molecules that currently have insufficient or non-existing spectral data. Therefore, is of great importance to obtain preliminary spectra that can lead to informed prioritization of molecules that have the most need for high quality spectral measurements or calculations.

For most molecules, however, approximate spectra will be sufficient. For example, most astronomical observations do not require high accuracy spectral data, full coverage or absolute completeness. On the other hand, low resolution data, particularly when combined with poor fundamental molecular data, can lead to misassignment of spectral features and consequently ambiguity in molecular detections.

RASCALL (Rapid Approximate Spectral Calculations for ALL) is a theoretical computational chemistry method that combines experimental measurements, organic chemistry, and quantum mechanics to provide approximate spectra for any given molecule. RASCALL has a multitude of potential applications, from astronomical insights into gaseous biomarkers, for which it was created, to industrial uses such as pharmacology and combustion analysis. RASCALL 1.0, presented here, primarily uses structural chemistry to simulate spectral data for thousands of molecules. RASCALL 1.0 relies on the spectral predictability of specific functional groups to estimate their collective contribution to an individual molecule's spectrum. The RASCALL 1.0 methodology builds on previous work described in \cite{atmos}.

The preliminary RASCALL 1.0 database currently contains 16,367 molecules and can be obtained online as the RASCALL 1.0 catalogue\cite{rascall_database}. However, we recommend that an updated version of the database be generated through the RASCALL code \cite{github}, or by contacting the corresponding author; see Appendix C for more details on the RASCALL code, and Appendix D for guidelines on using the  RASCALL molecular database.

The version of RASCALL presented here aims to address the following three concerns:
\begin{enumerate}
\item{Thousands of molecules have incomplete or completely absent spectral data, making any future detections of such molecules difficult or impossible. Through computational chemistry, RASCALL provides approximate spectra for most molecules, and some spectral information for all molecules. The associated preliminary database is publicly available and routinely updated\cite{github}. The current RASCALL repository\cite{rascall_database} contains spectral data for more molecules than any other database.}
\item{Experimental and theoretical methods for obtaining high accuracy spectra are extremely costly and time consuming. By providing preliminary spectra for thousands of molecules, and through the ability to analyse hundreds of molecules simultaneously, RASCALL provides an informative method of triage to guide, and prioritize, high accuracy studies.}
\item{Detections of molecules in atmospheric spectra are vulnerable to false positives, false negatives, and missassignments. RASCALL can highlight possible ambiguities in molecular detections and direct observations towards regions of any given spectrum that can help reduce the degeneracy in molecular identification.}
\end{enumerate}

We first provide a brief introduction to the concept of molecular functional groups and their spectral consequences. We then describe the RASCALL code and its functionality. Examples of RASCALL applications are then provided in the following section. Finally, the limitations of the RASCALL program, output types, and future developments are discussed.


\section{Molecules and their functional groups}\label{sec:functional}

Since the early 20th century, organic chemists have recorded molecular spectra and assigned features to their corresponding functional groups so that they could identify an unknown molecule by its component parts (see, for example, \cite{mcmurryOrganicChem}). Functional groups are defined as groups of atoms and bonds responsible for the specific physico-chemical properties of a particular compound. Each functional group has characteristic spectral features that remain mostly unchanged regardless of the compound in which they are present. The spectral manifestations of each functional group will depend mostly on its elemental composition, structure and bond types within the group, as well as its allowed symmetries of motion. Common functional groups include benzene rings, triple bonded carbons in alkynes, and the carbon-oxygen single bond of ethers. For the present work we have analyzed the most common functional groups composed of carbon, oxygen, nitrogen, sulphur, hydrogen, and phosphorus (CONSHP), as well as some common halogens. Future versions of RASCALL will be expanded to include more elements as long as they are able to form stable compounds.

For the creation of RASCALL 1.0 we analysed 120 functional groups associated with 19 different bond types, and constructed a table of the spectral properties for each of their spectrally active rovibrational symmetries. This table is compiled into a collection that we call the {\it{functional group dictionary}}. Whenever new functional groups are identified by the RASCALL team, these are added to the table alongside their predicted spectral properties.
Further details on the assessment of functional groups and the creation of the functional group dictionary can be found in Appendix A. Table \ref{tab:func}  shows a sample of the functional group dictionary. The RASCALL 1.0 functional group dictionary is available online as part of the catalogue documentation\cite{rascall_database}, and the most up-to-date table of functional group data is kept as a living document in the RASCALL GitHub repository\cite{github}.

\begin{table*}[]
\caption{\footnotesize{Sample of functional groups and their measured spectral properties. Information taken from \cite{irPDF, 05irPDF, ISAT}. The full functional group table is available online\cite{github,rascall_database}. Further details provided in Appendix A. \label{tab:func}
}}
\centering
{\small
\begin{tabular}{clccc}  
\hline
\hline
\noalign{\smallskip}
Functional Group Type & Symmetry & Location (cm$^{-1}$) & Strength & Additional information\\
\noalign{\smallskip}

\hline

\noalign{\smallskip}
=C-H & Stretching & 3010 - 3040 & Medium & Tri-alkyl Alkene\\
=C-H  & Bending & 950 - 1225 & Weak, narrow & Mono-alkyl Arene\\
C=O  & Stretching & 1720 - 1740 & Strong & Aldehyde \\
-CH$_3$& Bending & 1370 - 1390 & Strong & Alkane, deformation \\ 

S=O  & Stretch & 1380 - 1415& Strong & Sulfate\\ 
\hline
\vspace*{-5mm}
\end{tabular}
}

\vspace{1cm}
\caption{\footnotesize{Sample of the molecular dictionary: molecules, shown with IUPAC names and molecular formulas, and their associated functional groups. The number of times each functional group appears in the molecule is shown in brackets. The full molecular dictionary is available online\cite{github,rascall_database}. Further details provided in Appendix B.
 }}\label{tab:dic}
\begin{tabular}{cll}  
\hline
\hline
\noalign{\smallskip}
Molecule & Molecular Formula & Functional Groups (Incidence) \\
\noalign{\smallskip}

\hline

\noalign{\smallskip}
2-(ethylamino)-ethanol & C$_4$H$_{11}$NO & CH$_3$ (1), CH$_2$ (3), N-H (1), O-H (1) \\
methoxyphosphonic acid  & CH$_5$O$_4$P & CH$_3$ (1), -P-O-C (1), P=O (1)  \\
4,5-dihydro-1H-pyrrol-2-amine  & C$_4$H$_7$N$_2$  & CH$_2$ (2), =C-H (1), N-H (1),  NH$_2$ (1)\\
\hline
\end{tabular}
\vspace{0.5cm}
\end{table*}

Theoretically, a spectroscopically active molecule can be identified by its spectrum through the assignment of all of its features to their corresponding functional groups. Likewise, the spectrum of any molecule can be theoretically approximated by reverse engineering it from the collective spectral contributions of that molecule's functional group constituents.

Establishing which functional groups are present in an arbitrary molecule is not trivial. Although any molecule can be queried for the presence of a functional group, each functional group can vary widely in the effective number of bonds and atoms within it and each atom can be an active participant in several distinct functional groups. Additionally, functional groups within a molecule may not be spectrally active, or only extremely weakly; we call these {\it{muted functional groups}}.

For RASCALL 1.0, we assigned functional groups to a total of 16,367 molecules, selected from the repository of small volatile molecules published in \cite{seager2016toward}. We refer to the complete list of molecules and their associated functional groups as the {\it{molecular dictionary}} (see sample in Table \ref{tab:dic}). 
The molecular dictionary used in RASCALL 1.0 is available online as a supporting document to the preliminary RASCALL catalogue\cite{rascall_database}, and its most current version will be maintained in the RASCALL GitHub repository\cite{github}.

Further details on the types of molecules considered, and on the assignment of functional groups to molecules, can be found in the following section, and Appendices A and B. The Discussion section describes issues resulting from simulating molecular spectra using only functional group theory, such as perturbation to the spectral features of a functional group due to the different neighboring atoms and bonds in the molecule. The Discussion section also includes our plans to address these issues in future versions of RASCALL.


\section{RASCALL 1.0}\label{sec:code}

RASCALL 1.0 has three main capabilities: 
\begin{itemize}
\item{Simulating spectral data for any molecule by plotting the positions and relative intensities of the molecule's band centers, as determined by the known, and predicted, spectrally active functional groups within the molecule. When available, RASCALL 1.0 can also plot experimental molecular spectra for comparison with our simulated outputs.}
\item{Grouping molecules based on key parameters: feature strength, wavelength range, functional group and molecular family. RASCALL 1.0 is then able to filter molecules with specified characteristics and provide statistics on the spectral and structural properties of molecular groups.}
\item{Identifying potential degeneracies in molecular detections by listing molecules with the same functional groups and consequently likely to have similar spectral features. Molecules can be plotted to inspect spectral similarities, and ambiguities can be reduced by highlighting wavelength regions with distinguishing spectral features.}
\end{itemize}

To perform the above tasks RASCALL requires a molecular dictionary and a functional group dictionary as input. For some molecules, we modify the functional group data to create custom-made spectra that accounts for the interaction between functional groups and physical chemistry predictions (see Discussion section for more details). 
RASCALL then outputs spectral information for any chosen molecule. Every time new patterns of functional group behaviour are established, new modifications are integrated into the RASCALL code, and updated molecular databases are generated. All associated data and documentation are maintained online\cite{github}.
The  requirements, methodology and outputs of RASCALL 1.0 are outlined in the following section. Further details of the coding structure can be found in Appendix C.

The RASCALL code is available online\cite{github} and is maintained and updated by the RASCALL team. It will be continuously improved throughout the RASCALL project, and can be used to generate an updated database locally. See "Future Versions" Section (~\ref{sec:future}) for information on upcoming RASCALL improvements and Appendix C for guidelines on using the RASCALL 1.0 code.

\subsection{The RASCALL 1.0 code}

\begin{figure}[ht!]
\includegraphics[width=0.5\textwidth]{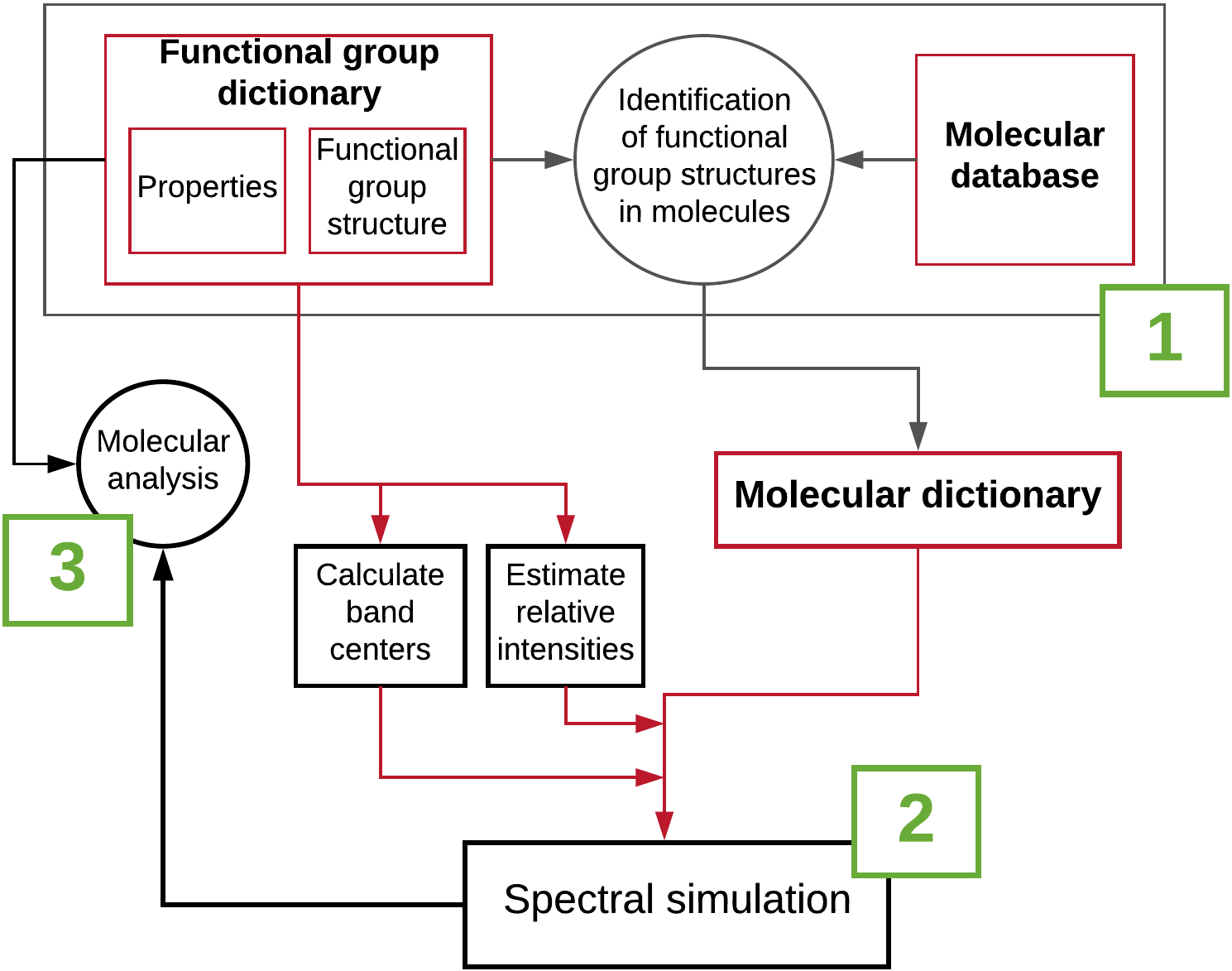}
\caption{\footnotesize{Flowchart outlining (1) the creation of the RASCALL 1.0 inputs (see Section \ref{inputs}), (2) the simulation of molecular spectra (see Section \ref{sec:simulation}), and (3) the analysis of the molecular outputs.\label{flowchart}}}
\end{figure}

RASCALL 1.0 is a group of python modules that can be considered as belonging to three connecting systems, outlined in Figure \ref{flowchart}: (1) input creation; (2) spectral simulation; and (3) molecular analysis. Input creation and spectral simulations are described in this section. Molecular analysis (3), is described throughout the remainder of the paper and includes spectral comparisons (between RASCALL 1.0 spectra and experimental data), and RASCALL 1.0 applications (see Applications, Section 4). RASCALL 1.0 and all its future versions can be used as a black-box program, but the inner components and logic are available online for use, inspection and modification\cite{github}.

\subsection{RASCALL 1.0 Inputs (1)}\label{inputs}

RASCALL 1.0 requires as core inputs a molecular dictionary and a functional group dictionary (see Figure \ref{flowchart}).

The functional group dictionary is created from a standardized table of the most common known functional groups and their spectral properties: symmetry type, frequency range, line shape, intensity and source (see Table \ref{tab:func}). The functional group table contains information collected from the literature, experimental data and theoretical inference. For each functional group that is identified, RASCALL collects every known and assigned fundamental vibrational mode (e.g., stretching, bending), as well as combination and hot bands, where these are available. The labeling of vibrational modes in the literature is often absent, or limited in information, so RASCALL 1.0 uses placeholder labels for modes that are spectrally active, and included in the RASCALL spectra, but not yet linked to specific motions within a molecule (e.g., torsional modes, higher excitations). Future versions of RASCALL will include more information on type of the motions responsible for the molecular features plotted in RASCALL spectra.

The functional group table is read into RASCALL and stored internally as a functional group dictionary. The raw functional group table will remain a living document, continually updated as new sources are uncovered and theoretical predictions enhance functional group information. For further details on the creation, validation and maintenance of the functional group table see Appendix A.

To create the molecular dictionary RASCALL identifies the functional groups present within any given set of molecules (see Table \ref{tab:dic}). RASCALL 1.0 uses the All Small Molecules (ASM) set of molecules \cite{seager2016toward}, which currently contains 16,367 members with $\leq$6 non-hydrogen atoms, that are volatile at standard temperatures and pressures, and primarily composed of the most common elements, CONSHP (with some additional exceptions, such as common halogens). See \cite{seager2016toward} for further information on the composition and creation of the ASM set of volatile molecules. 

To identify which functional groups are present in each ASM molecule, we use a custom python code, based on the rdkit chemoinformatics sub-structure matching module \cite{landrum2006rdkit}. The matching module was written to identify which spectroscopically active functional groups are present, and how many times, in which molecules (examples listed in Tables \ref{tab:func} and \ref{tab:dic}). The resulting list of molecules and their constituent functional groups is then stored internally in RASCALL as a molecular dictionary.

The creation of the molecular dictionary is a difficult process, as the algorithm must be able to distinguish between functional groups in a molecule, without over- or under-identifying functional groups. Aditionally, each atom within a molecule can contribute to multiple functional groups, depending on its neighboring atoms and associated bonds. The RASCALL module for the identification of molecular functional groups correctly identifies unique and repeated functional groups, without falsely identifying groups that have no spectral consequences.  Further details into developing an accurate and exhaustive identification of all spectroscopically active functional groups within a molecule are provided in Appendix B.

Finally, RASCALL 1.0 takes as input experimental spectra from the NIST and the PNNL databases  \cite{nist,johnson2004pnnl} for comparison to the RASCALL 1.0 simulated spectra.

\subsection{RASCALL 1.0 Outputs}\label{sec:outputs}

The main output of RASCALL 1.0 is approximate spectral data for molecules in the gas phase. RASCALL draws on a library of functional group data and their associated properties to simulate the main spectral features of any molecule for which individual functional groups can be identified. Together these features create an approximate spectrum, which can be calculated by RASCALL 1.0 in seconds. The resulting database, currently containing 16,367 molecules, forms the RASCALL catalogue \cite{rascall_database}. We note that the RASCALL 1.0 database is large but it is not complete; although some molecules have reliable data on every one of their major spectral features, some molecular entries remain incomplete, or incorrect. As the RASCALL team addresses these issues, the RASCALL catalogue of molecules is continually improved and updated. The RASCALL 1.0 catalogue is easily downloaded\cite{rascall_database}, and contains spectral data for every molecule in the database as well as information about the functional groups identified in the molecule. Updated versions of the RASCALL database can be generated at any time through the RASCALL code\cite{github}, or by contacting the corresponding author.  For more information on the RASCALL database see Appendix D.

In addition to spectral data, the analysis modules on RASCALL 1.0 can provide the following outputs: information on the types and strength of functional groups in any given wavelength window; lists of molecules containing specific functional groups; number of predicted, spectrally active, functional groups and vibrational symmetries in any given molecule; statistics on the most common functional groups, or their properties (e.g. bond type, number of atoms); frequency of experimental measurements for different functional groups (based on spectral measurements of their molecules). For a list of future RASCALL output types and functionality see Section \ref{sec:future}.

\subsubsection{Spectral Simulation (2)}\label{sec:simulation}

\begin{figure*}[ht!]
\centering
\includegraphics[width=\textwidth]{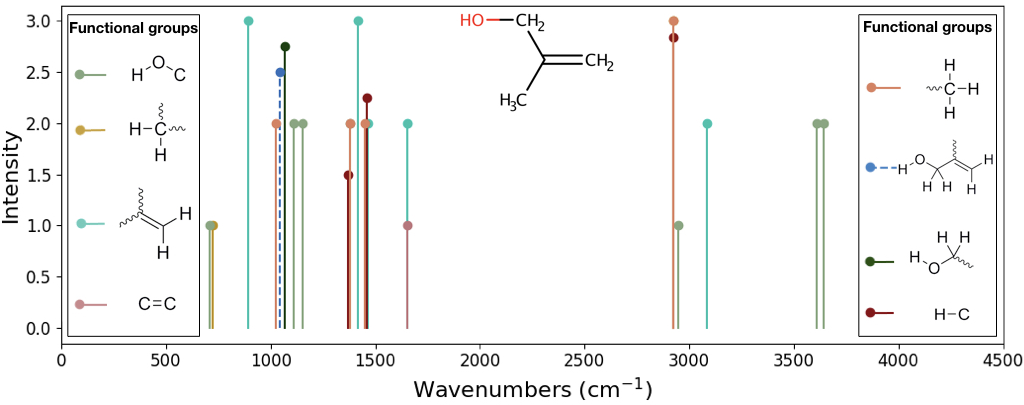}

\caption{\footnotesize{Example of the RASCALL 1.0 skeleton spectrum for the isopropenyl carbinol molecule (C$_4$H$_8$O). Structure of the molecule is shown in the center of the figure. Molecule is composed of 8 unique functional groups with a total of 24 symmetries between them. Functional group structures are shown on either side of the plotted spectrum. Straight lines represent single bonds to between represented atoms, or to a carbon atom if no letter is shown. The position of the vertical lines represents the approximate frequency of the molecule's band centers (in wavenumbers), and their height represents the bands' qualitative intensity, with values ranging from 1 to 3 (representing weak to strong absorptions, respectively). All symmetries belonging to the same functional group are plotted in the same color. Functional groups predicted by the RASCALL team are plotted with dashed lines.}\label{skeleton}}
\end{figure*}

The core RASCALL modules use the data in the molecular and functional group dictionaries, as well as any predicted enhancements to those data, to reverse engineer an approximate spectrum for any molecule in the ASM database. In RASCALL 1.0, this spectral simulation is done by processing the properties of the functional groups within a molecule and plotting them into a "skeleton" spectrum composed of the qualitative intensities and approximate central positions of a molecule's most common functional groups (see Figure \ref{skeleton}). 

All vibrational modes of a given functional group are plotted in the same color (e.g., the symmetric stretch and bending modes of CH$_2$). The dashed lines represent new functional groups predicted by RASCALL, or its precursor ATMOS\cite{atmos}. For RASCALL 1.0, these new functional groups are obtained in two ways. By extracting the spectral information from experimental spectra (e.g. NIST data \cite{nist}) and by theoretically extrapolating from existing functional group data. The latter approach relies on the RASCALL team's predictions for the spectral behavior of new functional groups with similarities to known functional groups. We name these new RASCALL functional groups {\it{associated functional groups}}.

Functional group sources in the literature provide limited data on experimental band shapes so RASCALL 1.0 uses approximate band centers as vertical lines with zero-width. The positions of the band centers are calculated by taking an average frequency from the range of positions reported in the literature and in experimental databases. RASCALL also collects information on band shapes and widths where these are available; future RASCALL spectra will reflect these properties.
 
The literature on functional group intensities only provides qualitative values, so RASCALL 1.0 divides functional group features into relative strengths, ranging from "weak" to "strong", which are then translated into qualitative intensities ranging from 1 to 3. Future versions of RASCALL will add quantitative intensities to the database (see Section \ref{sec:future}), by extracting absolute values from reliable experimental measurements and theoretical calculations where available, and from approximate quantum chemistry simulations where no reliable data exists.

In a few cases, RASCALL intensities are reduced from those reported in the literature to account for muted functional groups (see Discussion for further details). In general, where the RASCALL team finds that the intensities (or frequencies) of a functional group mode are consistently over- or under-estimated in the literature, the functional group table is updated. Whenever this occurs, new spectral data for all molecules in the RASCALL database with the affected functional group are generated. 
\subsubsection{Spectral Comparison}\label{sec:comparison}

\begin{figure*}[ht!]
\centering
\includegraphics[width=\textwidth]{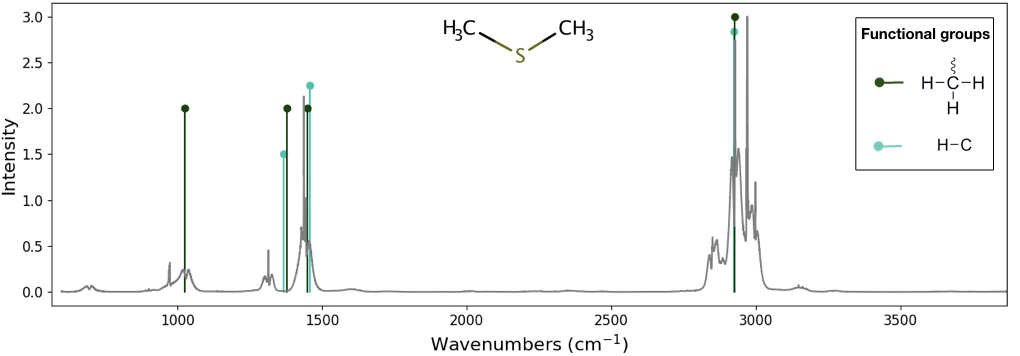}
\caption{\footnotesize{Comparison between the experimental spectrum  for the dimethyl sulphide molecule (C$_2$H$_6$S) and its theoretical skeleton spectrum from RASCALL 1.0. Theoretical skeleton spectrum (vertical lines) is created by RASCALL 1.0, with line position and height representing the approximate frequency and qualitative intensity, respectively, of the molecule's band centers. Experimental data (gray continuous line) is obtained from the PNNL database \cite{johnson2004pnnl} with relative absorbance scaled to the RASCALL 1.0 intensities. Structure of the molecule shown in the center of the figure. Functional group structures are shown on right side of the plotted spectrum. RASCALL 1.0 recognises two spectrally active functional groups (C-H and $-$CH$_3$) in the dimethyl sulphide molecule, between them containing seven vibrational modes (e.g., bending and stretching motions). RASCALL identifies that  there two -CH$_{3}$ groups present in the dimethyl sulphide molecule, which result in the strong double peaks in the 2750 - 3100 cm$^{-1}$ region, but RASCAL 1.0 cannot distinguish their spectral contributions and predicts a single peak. Future RASCALL 1 updates will address this issue.}
\label{mol_example}}

\end{figure*}

\begin{figure*}[ht!]
\centering
\includegraphics[width=\textwidth]{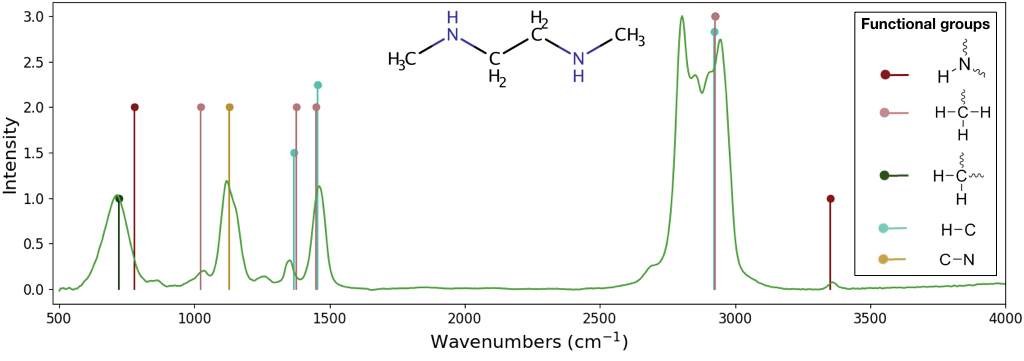}
\caption{\footnotesize{Comparison between the experimental spectrum for the C$_4$H$_{12}$N$_2$ molecule and its theoretical skeleton spectrum from RASCALL 1.0. Theoretical skeleton spectrum (vertical lines) is created by RASCALL 1.0, with line position and height representing the approximate frequency and relative intensity, respectively, of the molecule's band centers. Experimental data (green continuous line) is obtained from the NIST spectral database \cite{nist} with relative absorbance scaled to the RASCALL 1.0 intensities. Structure of the molecule shown in the center of the figure. Functional group structures are shown on right side of the plotted spectrum. RASCALL identifies five functional groups within the C$_4$H$_{12}$N$_2$ molecule, between them containing twelve spectrally active modes (e.g. stretching and bending vibrational motions of the methyl, -CH$_{2,3}$, and amine, -NH$_{1,2}$, groups). RASCALL 1.0 correctly predicts the approximate positions of the majority of the bands in the C$_4$H$_{12}$N$_2$ molecule and identifies its strongest and weakest features. RASCALL identifies that there are multiple incidences of the -CH$_{3}$ group, which result in the strong double peaks in the 2750 - 3100 cm$^{-1}$ region, but RASCAL 1.0 considers their spectral contributions indistinguishable and predicts a single peak. Future versions of RASCALL will be able to estimate the spectral consequences of repeated functional groups in a molecule. RASCALL 1.0 correctly predicts that the secondary amine group has one very weak feature at 3352 wavenumbers (in crimson) but incorrectly predicts a medium intensity feature at 780 wavenumbers. With RASCALL 2.0 it will be possible to determine whether the feature's strength was overestimated, whether the symmetry is not active in the C$_4$H$_{12}$N$_2$ molecule, or whether this feature's predicted position requires adjustment.}
\label{mol2_example}}

\end{figure*}

The RASCALL 1.0 spectra provides the approximate position and qualitative strengths of the band centers of any molecule in the database. Given an individual molecule or group of molecules, RASCALL can also provide information on whether there are any alternative spectral sources available in the literature for those molecules (e.g., ExoMol \cite{tennyson2016exomol}, HITRAN \cite{hitran2016}, NIST \cite{nist}, PNNL \cite{johnson2004pnnl}). Where it is available, RASCALL is able to immediately compare the RASCALL-simulated spectra to existing NIST and, in some cases, PNNL cross-sections. For this comparison, RASCALL 1.0 scales the relative absorbance of the cross-sections to the RASCALL 1.0 intensities.
As examples, Figures \ref{mol_example} and \ref{mol2_example} compare the simulated RASCALL 1.0 spectra for the C$_2$H$_6$S (dimethyl sulphide) and C$_4$H$_{12}$N$_2$ molecules with their experimentally measured spectra, as provided by the PNNL and the NIST databases, respectively.

As can be seen by the agreement with experimental measurements (see, e.g., Figures \ref{mol_example} and \ref{mol2_example}), RASCALL 1.0 can accurately predict the strength and position of the majority of the C$_2$H$_6$S (dimethyl sulphide) and C$_4$H$_{12}$N$_2$ features. RASCALL 1.0 correctly identified the strongest features in both molecules and highlighted areas of spectral complexity. We note that, in both molecules, there are duplicated methyl components (CH$_3$), that result in the strong double peaks between 2750 - 3100 cm$^{-1}$. RASCALL 1.0 considers any repeated functional group spectrally indistinguishable from a single incidence of that functional group, and consequently only one spectral peak is predicted when in reality the molecular band is split into two peaks. Future versions of RASCALL will address these {\it{degenerate functional groups}} and their spectral consequences (see Section \ref{future1}).

RASCALL 1.0 correctly assessed which atoms and bonds belonged to spectrally active functional groups, and which do not strongly contribute to the molecule's spectrum. For example, when simulating the dimethyl sulphide molecule, RASCALL 1.0 does not identify functional groups associated with the sulfur atom but the main spectral features of the molecule are still correctly predicted (see Figure \ref{mol_example}). RASCALL 1.0 recognizes that the spectrum of the dimethyl sulphide molecule does not depend strongly on the central atom (sulfur). In contrast, RASCALL identifies that the spectra of other dimethyl molecules, such as dimethyl oxide, dimethylmethane and dimethylamine, depends strongly on the central atoms (oxygen, carbon and nitrogen, respectively).
 
 In many cases, for example the small feature in the C$_4$H$_{12}$N$_2$ molecule at 3352 wavenumbers (see Figure \ref{mol2_example}), even extremely weak spectral features were predicted correctly and assigned the lowest intensity in RASCALL 1.0 ("weak", corresponding to a value of 1).
 
 The positions of the functional group features predicted by RASCALL 1.0 remain the same independently of which molecule they are in, despite there being variations between these in reality (see, for example, the prediction mis-match around 1450 cm$^{-1}$ for dimethyl sulphide in Figure \ref{mol_example}). These variations in the spectral behaviour of a functional group occur because each functional group does not exist in isolation and experiences perturbations from its structural context within the molecule. See Future Versions section for details on how we will address variations of functional group behavior.

All examples of approximate skeleton spectra shown here are produced by RASCALL 1.0 in a few seconds. RASCALL 1.0 is capable of producing equivalent spectra for any other molecule within a comparable timescale. The collection of spectral data for all 16,367 RASCALL molecules forms the RASCALL catalogue. We note that the RASCALL 1.0 catalogue of molecular data is extensive, but not complete. It is a preliminary database generated from the RASCALL code at the time of publication, which is only as complete as the molecular dictionary and functional group table used as inputs (see Section \ref{sec:discussion} for more details on the shortcomings of RASCALL 1.0). Some entries in the RASCALL 1.0 database have good spectral predictions for all major features of a molecule, but some molecules have missing features, or inaccurate predictions. The RASCALL catalogue is a living document that contains skeleton spectra and functional group information for all molecules in the RASCALL database. As the functional group data is expanded and enhanced, and the RASCALL code continues to develop, the database is updated. The current RASCALL 1.0 catalogue is available for download\cite{rascall_database}. Updated versions of the RASCALL database can be generated through the RASCALL code\cite{github} (see Appendix D for more details).


\section{Applications}\label{sec:applications}

\begin{figure}[ht!]
\centering
\includegraphics[width=0.5\textwidth]{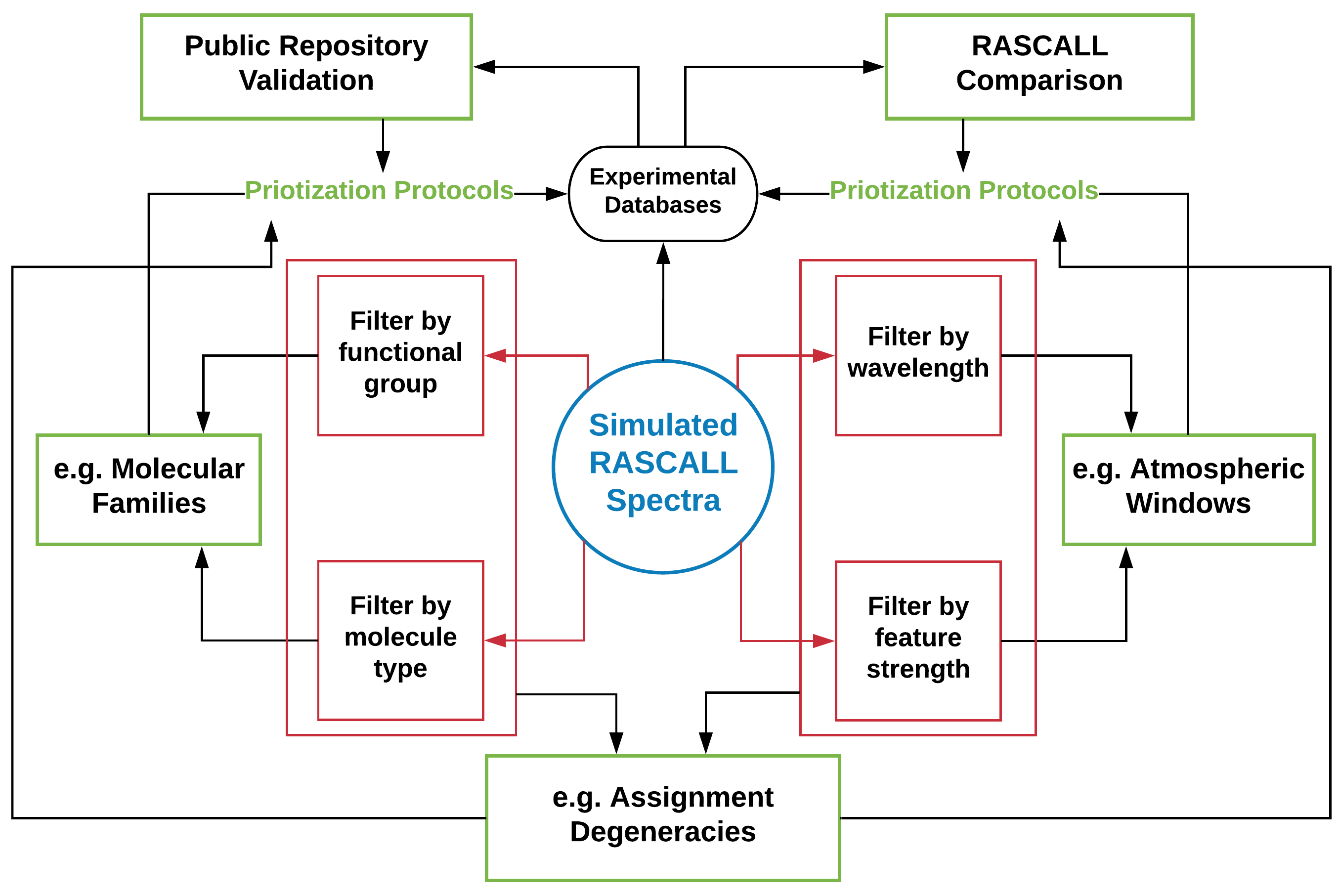}
\caption{\footnotesize{Flowchart outlining the analysis framework of RASCALL 1.0 (red boxes), and its associated applications (green boxes). The primary application of RASCALL 1.0 is as a simulator of approximate spectra (blue circle). Through comparisons to data in experimental databases (top-right), RASCALL 1.0 spectra can be analysed for accuracy and in turn validate the spectral information in public repositories (top-left). RASCALL 1.0 can perform statistical and spectral analyses on families of molecules through molecular type and functional group filtering (left panels). RASCALL 1.0 can also filter large sets of molecules by absorption region and feature strength which can be used to, for example, highlight promising molecules for detection in a variety of atmospheric windows (right panels). Finally, by making use of all filtering functions, RASCALL 1.0 can highlight possible degeneracies between molecules and direct observations to wavelength regions that can reduce ambiguity in molecular detection (bottom panel). Analyses from all of the previously mentioned applications can be used to inform prioritization protocols for future spectral measurements.\label{applications}}}

\end{figure}

RASCALL 1.0 is intended for use as a simulator of approximate spectral data for any molecule for which there are insufficient or no data available, which is the case for thousands of small volatile molecules. The primary application for RASCALL 1.0 is, therefore, as a producer of molecular data for input in atmospheric models where no other source (with higher accuracy) is available. However, even in its first iteration, RASCALL 1.0 has many additional applications, examples of which are outlined in this section and in Figure \ref{applications}.

\subsection{Public Repository Validation}\label{sec:validation}

RASCALL spectra and associated metadata can be used to validate spectral information in public repositories of molecular data, such as NIST \cite{nist} and other experimental databases. 

The spectral data in NIST originates from a wide variety of sources, which in turn have a wide range of reliability. NIST data often have a singular source, with limited information on the original experimental laboratory procedures and set-ups (e.g., sample purity). For hundreds of molecules, NIST is the only publicly available database containing infrared spectra. Without an alternative source to validate NIST data, mistakes go unnoticed for decades. With RASCALL we have been able to compare every single IR spectrum available in NIST (for molecules with $\leq$6 non-hydrogen atoms) for a least a first pass analysis. 

Spectra in experimental databases can be used as validation tools for the spectral outputs produced by RASCALL. In turn, by comparing the RASCALL output with experimental spectra in public repositories, and statistically analyzing discrepancies between spectra, it is possible to uncover mistakes and inconsistencies in these experimental databases. In the case of NIST, multiple issues were discovered through comparisons with RASCALL 1.0 spectra, such as: liquid/solid state spectra mislabelled as gas phase; unreliable, hand-drawn spectra legitimised through digitalization; undated, unlabelled spectra in the form of a scanned image (e.g., HCN); incorrectly assigned vibrational modes; inconsistent labelling between chemical formulas, IUPAC names and spectral data (e.g. 2,5-diazahexane labelled as C$_4$H$_{10}$N$_2$ instead of the correct C$_4$H$_{12}$N$_2$).

\begin{figure*}[ht!]
\centering
\includegraphics[width=0.75\textwidth]{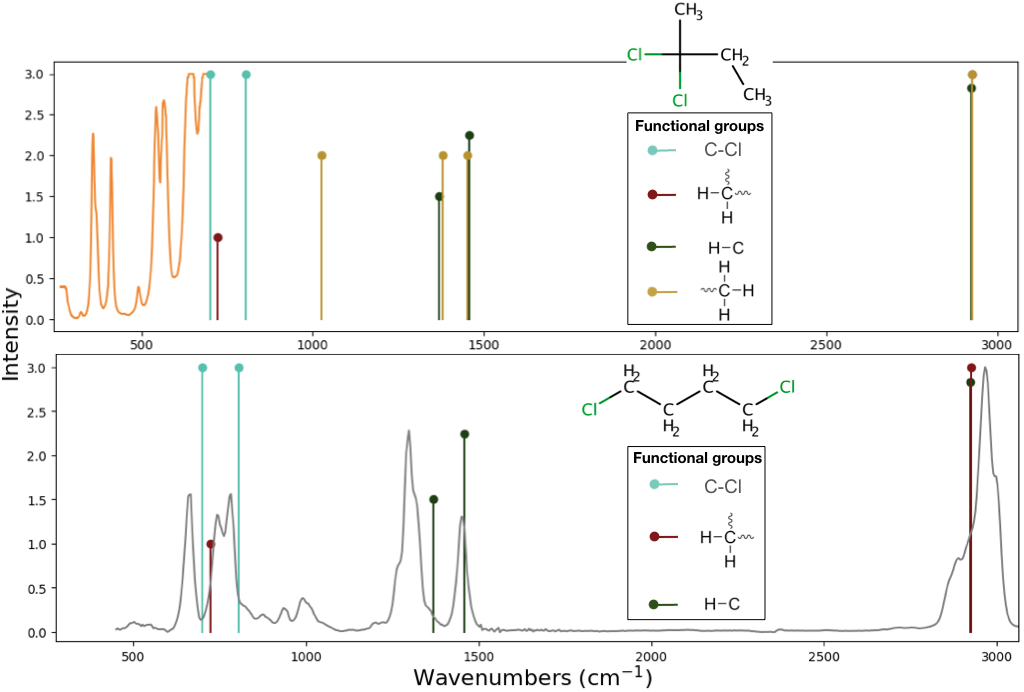}
\caption{\footnotesize{Comparison of the RASCALL 1.0 spectra (vertical lines) and the NIST spectra for two isomers of the C$_4$H$_8$Cl$_2$ molecule. Structures of the molecules and functional group labels shown around 2250 cm$^{-1}$. RASCALL spectra agrees with the most recent NIST data for 1,4 dichlorobutane (lower panel, continuous gray line) and is in strong disagreement with the older 2,2 dichlorobutane (upper panel, continuous orange line) NIST spectrum. We hypothesize that the data for 2,2 dichlorobutane is not from a gas phase measurement and should be labelled accordingly. 
RASCALL identifies three common functional groups in both isomers and one additional functional group in 2,2 dichlorobutane, associated with the CH$_3$ group. \label{dichloro}}}

\end{figure*}

Another example of a spectral issue found in the NIST database concerns two of the C$_4$H$_8$Cl$_2$ isomers: 1,4 and 2,2 dichlorobutane. Both isomers have unlabelled spectra in NIST that are in complete disagreement with RASCALL spectra. The original data was hand-drawn sometime prior to 1970, and subsequently digitalised and included in the NIST database. For the 1,4 dichlorobutane isomer, and additional recent gas phase spectrum is available in NIST, and this spectrum is in good agreement with the RASCALL predictions (see Figure \ref{dichloro}). In the case of 2,2 dichlorobutane, complete disagreement with RASCALL spectra strengthens our hypothesis that the NIST spectra is incorrect, or at least not measured in the gas phase.

Throughout its development cycle, RASCALL will continue to recommend corrections to the data in public repositories of molecular spectra, and database managers can use RASCALL to independently validate their data. 

\subsection{Molecular Prioritization Protocols}\label{sec:triage}

High-resolution experimental measurements and theoretical calculations of molecular spectra are difficult processes. Both require expensive equipment and very often several years of work even for small molecules; as an example, the calculation of the full spectrum for PH$_3$ required over three years of simulations and high-performance computing \cite{sousa2014exomol}. RASCALL can efficiently provide approximate spectra and simultaneously triage a large number of molecules, assessing their suitability for in-depth theoretical or experimental spectroscopic studies. Due to its accessibility and speed, RASCALL is an excellent tool to triage large groups of molecules and inform future prioritization for high resolution measurements and calculations of molecular spectra.

High resolution spectroscopy databases, such as HITRAN \cite{hitran2016} and ExoMol \cite{tennyson2016exomol}, can use RASCALL to scope potential molecules for in-depth treatment, to triage a large group of molecules when deciding which to prioritize, and to provide intermediate spectra while accurate models are being developed. 

As an example: In-depth studies of comet spectra require knowledge of the spectral behaviour of halocarbons \cite{fayolle2017protostellar}, which in turn require a comprehensive understanding of the common and unique features of each halocarbon of potential interest. RASCALL 1.0 can provide approximate spectra for 4067 halocarbons that are volatile in a wide range of temperature and pressure regimes. Out of those 4067 halocarbon molecules, RASCALL finds that less than 4$\%$ (144) have any spectral information available in the literature. Practically, it is important to prioritize those halocarbons that can benefit the most from high-resolution measurements. Ideally, future lab studies should work towards obtaining measurements for the missing spectra of these 3,923 halocarbons. In the meantime, the RASCALL catalogue provides preliminary data on all of the 3,923 halocarbon molecules that currently have absent spectra, and supplementary data for the 4067 halocarbons with alternative sources of spectra.

\begin{figure}[ht!]
\centering
\includegraphics[width=0.5\textwidth]{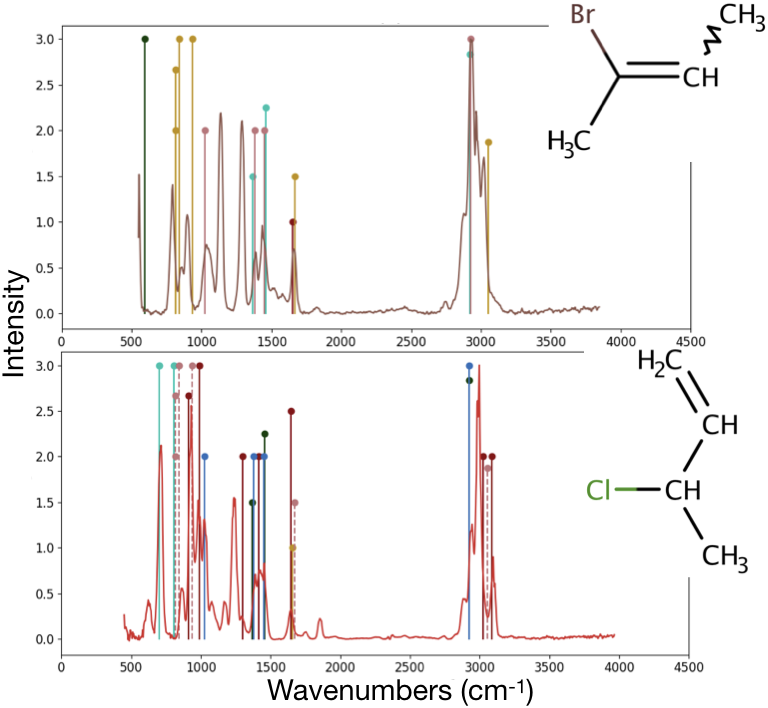}
\caption{\footnotesize{Comparison of the NIST\cite{nist} spectra (continuous lines) to the RASCALL spectra (vertical lines) for the 2-Bromo-2-butene (top panel) and $\alpha$-Methallyl chloride (bottom panel) molecules. Structures of the molecules shown to the right of their spectra. Both RASCALL 1.0 and NIST spectra show that the two molecules' strongest feature is caused by the -CH$_3$ feature around 3000 wavenumbers. Low resolution measurements may not be sufficient to distinguish between the fingerprint region (450 - 1600 wavenumbers) of the two molecules. \label{halo}}}

\end{figure}

The detection of any halocarbon requires some knowledge of its spectra, but the unambiguous identification of halocarbons that have similar functional groups requires high-resolution spectra. RASCALL can point out which molecules have similar functional groups within a family of molecule, and therefore should be prioritised for future high-accuracy studies. For example, RASCALL shows that a large number of halocarbons have many common functional groups, such as the -CH$_3$ functional group, and as such display similar spectral behaviour. In fact, out of the 144 halocarbons that have spectra in the NIST database \cite{nist}, 58 contain the -CH$_3$ group, and in 32 of those the -CH$_3$ group is the cause of the halocarbon's strongest spectral feature (see, for example, 2-Bromo-2-butene and $\alpha$-Methallyl chloride, shown Figure \ref{halo}). RASCALL can  point out which members of the halocarbon family, and any other molecular group, have most spectral similarities between them and are therefore in most need of detailed follow-up.

\subsection{Filtering by Wavelength Region}\label{sec:windows}

In most atmospheric spectra there will be wavelength regions that are oversubscribed by the atmosphere's most dominant molecules. Any minor gas that might be contributing to an atmosphere will be difficult to detect unless it absorbs in wavelength regions where no other major molecule is absorbing; we call these unpopulated wavelength regions {\it{atmospheric windows}}. Knowing which molecules are spectrally active in any given atmospheric window allows us to be focused in our search for every possible atmospheric component of interest.

RASCALL 1.0 can analyze the approximate spectra for thousands of molecules and identify which have significant spectral features in the atmospheric windows of different environments. Given a set of atmospheric windows where non-dominant molecules can be detected, RASCALL 1.0 can quickly determine which gases would exhibit strong features in those regions of an atmospheric spectrum.  RASCALL 1.0 allows the user to choose either a custom wavelength window or a set of windows within a preset atmosphere (e.g. Earth-like, pre-biotic Earth-like, methane-rich). RASCALL 1.0 can also identify which of these molecules have available spectra in the literature and provide approximate spectra for those molecules where there is none. Molecules for which the RASCALL 1.0 spectra will not suffice can be recommended as target molecules for in-depth experimental or theoretical studies.

As an example: future ground-based telescopes (e.g., ELT\cite{tamai2014european}, TMT\cite{skidmore2015thirty}) will have powerful observing capabilities but will remain limited by the atmospheric windows of the Earth. Our atmosphere is partially transparent in a few near and mid-infrared regions, such as the L (2710.0 to 3115.3 cm$^{-1}$), M (2008.0 to 2207.5 cm$^{-1}$) and N (851.1 to 1081.1 cm$^{-1}$) windows. Out of a total of 16,367 volatile molecules that are analysed by RASCALL 1.0, 15,862 are absorbers in at least one of the L, M or N windows. The L window (2710.0 to 3115.3 cm$^{-1}$) is particularly rich in molecular spectra because it contains one of the strongest symmetries of the C-H functional group. There are 919 molecules that can have strong features in the features in the M window (2008.0 to 2207.5 cm$^{-1}$) and 4,216 in the N window. These are large numbers of molecules but provide an unbiased starting point for the investigation of which molecules could cause unknown spectral features in L, M or N windows. Furthermore, these sets of molecules can be narrowed down to account for context (e.g., atmospheric cycles, reactivity rates, destruction mechanisms).

RASCALL 1.0 allows for fast filtering of lists of molecules by frequency, intensity, structure and molecular family, i.e. according to the types of atoms, functional groups, and spectral properties of each molecule. As such, RASCALL 1.0 is capable of high-throughput predictions that allow for a big-picture analysis of the molecular detection potential of a given spectral window. This in turn allows for a rough estimation of a number of potential molecules that could in principle occupy a given spectral window. Consequently, RASCALL predictions of potential "spectral window crowding" provide a theoretical estimate of a degree of degeneracy in the observational data. An example of such an analysis include identifying that a single atmospheric window may be insufficient to distinguish between two or more crucial molecules that are targets for detection. In these cases, the big data approach to molecular assignments that RASCALL takes can provide an assessment of potential degeneracies and direct future measurements to wavelength regions that can help resolve any ambiguities. 

Currently RASCALL 1.0 is being used to select molecular candidates that can be responsible for unknown IR features in observed astronomical spectra (e.g., from the warm Neptune GJ 436b\cite{lothringer2018hst}), based on the strength and location of the features as well as the expected temperature, pressure and chemical network in the observed atmospheres. Due to the limited temperature and wavelength range of RASCALL 1.o, only IR observations of warm, or cool, astronomical bodies can be analysed. Future versions of RASCALL will cover a wider range of temperatures and wavelengths, allowing for the analysis of unidentified spectral features beyond the IR (e.g., HAT-P-26b \cite{wakeford2017hat}), and from hotter astronomical bodies (e.g., WASP-121b \cite{evans2018ultrahot, evans2017ultrahot}).

\subsection{Molecular Families}\label{sec:families}

Any industries or research areas where the remote detection of unknown gases is required can benefit from using RASCALL. Examples of this would be the pharmacological research aiming to identify impurities in chemicals, and the oil industry where RASCALL could help monitor distillation processes by distinguishing between volatiles being produced. RASCALL can analyze thousands of molecules and group them by composition, bond types, and functional groups. Given a selection of molecules that are to be expected in a particular environment, RASCALL can provide similarities and distinguishing features between all molecular candidates.

Many organic molecules contain very common and often structurally similar functional groups. It is a non-trivial task to comprehensively understand what structural features of individual molecules make them spectrally unique and distinguishable from other molecules with the same functional groups. This unappreciated problem is particularly common within families of compounds that contain very common functional groups, such as the C$=$C, CH$_3$ and CH$_2$ groups, which are present in thousands of different molecules. 

One example of a molecular family with a large number of common functional groups is the hydrocarbon family. Hydrocarbon-rich environments, which are common both in industry (e.g. petroleum combustion, catalytic cracking) and in astrochemistry (e.g. the atmosphere of Saturn's moon, Titan), have complex atmospheric spectra with contributions from many hydrocarbons (e.g., C$_2$H$_6$, C$_2$H$_4$ and C$_2$H$_2$).

\begin{figure}[ht!]
\centering
\includegraphics[width=0.5\textwidth]{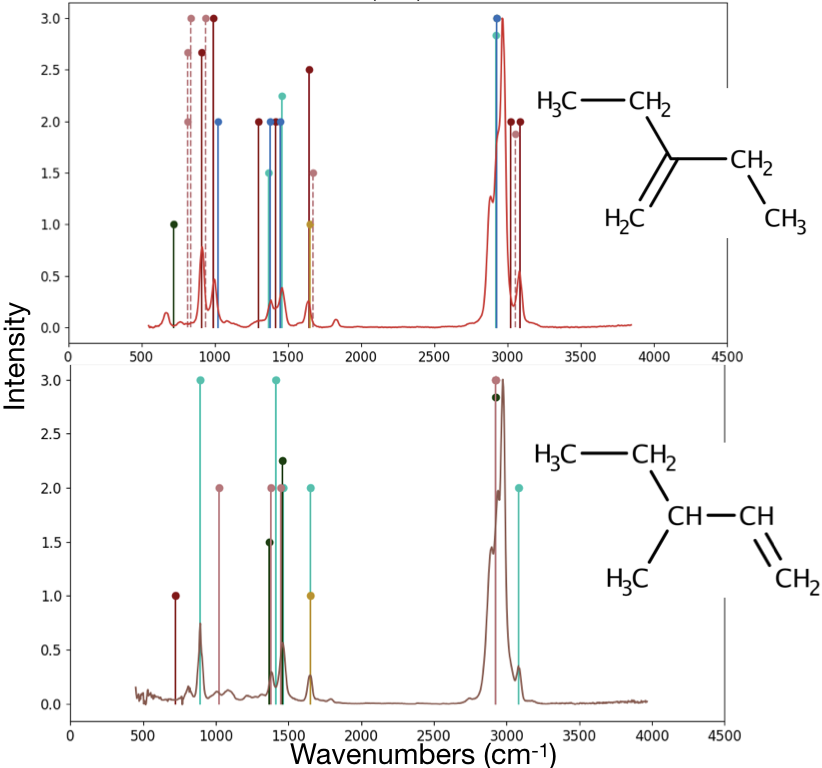}
\caption{\footnotesize{Comparison of the NIST\cite{nist} spectra (continuous lines) to the RASCALL 1.0 spectra (vertical lines) for 3-Methyl-1-pentene (top panel) and 1-butene (bottom panel). Structures of the molecules shown to the right of their spectra. Most features of both molecules are predicted by RASCALL 1.0. RASCALL 1.0 can produce approximate spectra of equivalent quality for any other 408 hydrocarbon molecules in the RASCALL database.\label{hydro}}}

\end{figure}

RASCALL 1.0 identifies 408 different hydrocarbons in the current database. All are small hydrocarbons (less than 6 non-hydrogen atoms) that are volatile in a wide range of temperature and pressure regimes. Out of those 408 hydrocarbons, less than 19$\%$ (76) have any spectral information available in the literature (e.g., NIST \cite{nist}, HITRAN \cite{hitran2016}). The remaining  332 hydrocarbon gases have completely unknown spectra and as such are impossible to identify even if present in high abundances in a hydrocarbon rich atmosphere. For an example of the importance of access to a wide range of spectra for hydrocarbon molecules in the gas phase, one can consider the process of fluid catalytic cracking (FCC) \cite{lopez2018monitoring}. The analysis of the reaction process can only be performed by detecting specific hydrocarbons in the IR, with molecular species arising throughout the conversion process. Currently spectral data for many hydrocarbon species relevant for FCC are missing, which limits the ability to predict the fluid dynamics and cracking kinetics of the FCC process.

RASCALL 1.0 can reproduce the strongest features of the hydrocarbons for which there is experimental data available (see, for example, Figure \ref{hydro}) and provides approximate spectra for the 81$\%$ of the hydrocarbons for which there is currently no spectra available at all.
These approximate spectra can both help with detecting molecules in mixed spectra and provide side-by-side comparisons of any hydrocarbons to highlight possible regions of degeneracy and distinguishability.
												
As with previously mentioned applications, the insights obtained from RASCALL with regards to hydrocarbon structural diversity and any other possible degeneracies can be used to inform prioritization protocols for high accuracy studies.

\subsection{Assignment Degeneracies}\label{sec:ambiguities}

Molecular identifications are often made by assigning one, or a small group, of spectral features to molecular candidates. However, at all but extremely high resolutions, most individual features could belong to a large number of molecules.

The remote detection of molecules within an environment requires that molecules' features be resolved and identified unambiguously. Given that many molecules have similar features and most recorded spectra has many sources of noise and uncertainties, the assignment of individual or a collection of features must be done carefully. RASCALL can identify which molecules have spectral traits that are compatible with any given observed features, therefore highlighting possible degeneracies in assignment. RASCALL can also provide a wider range of features for each potentially detected molecule so that the ambiguity in any identification can be reduced. We note that ambiguity in the spectral identification of molecules can also be mitigated through understanding the chemical network of the observed target (which would reduce the number of plausible molecular candidates for each feature) and by considering which molecules are likely to be stable in the temperature and pressure regimes considered (e.g., few polyatomic molecules will survive stellar and hot-Jupiter environments).

\begin{figure}[ht!]
\centering
\includegraphics[width=0.5\textwidth]{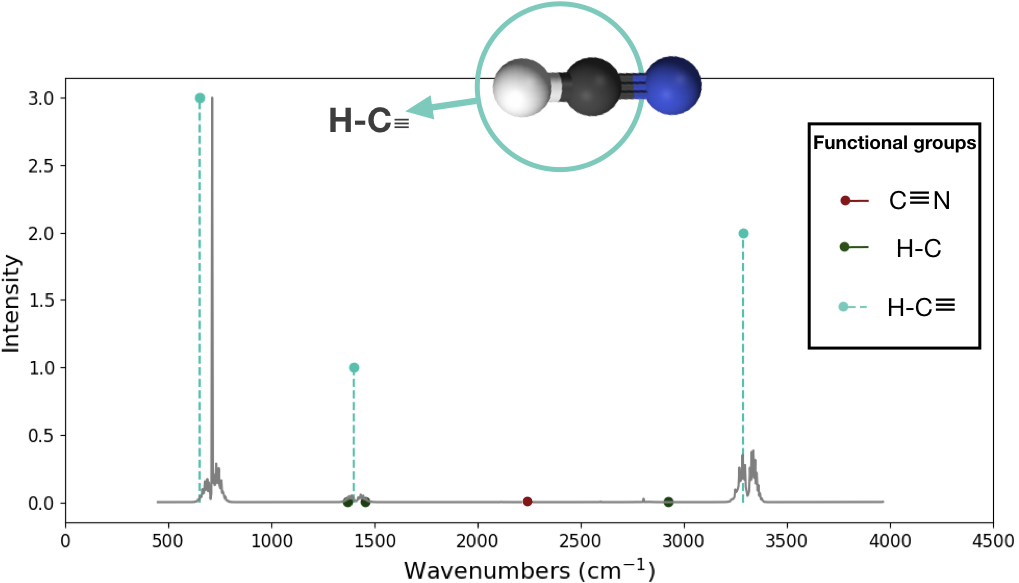}
\caption{\footnotesize{Comparison of the ExoMol\cite{tennyson2016exomol} spectra (continuous gray line) to the RASCALL 1.0 spectra (vertical lines) for the hydrogen cyanide (HCN) molecule. Only one of the three functional groups in the HCN molecule is spectrally active:  $\equiv$C-H. The $\equiv$C-H functional group is illustrated at the top of the figure in turquoise, with associated spectral features as dashed lines. The other two functional groups, C$\equiv$N and H-C, are muted in the HCN molecule (see Discussion section for more details on muted functional groups). \label{hcn}}}

\end{figure}

As an example, one can consider hydrogen cyanide, or HCN. HCN is presumed to be common in any hydrocarbon-rich environment, such as the atmosphere of Titan \cite{marten2002new} or in petroleum refineries \cite{zhao1997nitrogen}, but HCN may also have been a crucial precursor to the origin of life \cite{ferris1972chemical,matthews2004hcn}. The main features of the infrared spectrum of HCN can be completely explained by its functional group $\equiv$C-H, which contains a carbon with a single hydrogen and a triple bond to nitrogen (see Figure \ref{hcn}). The $\equiv$C-H functional group alone, through its fundamental bending motion (strong feature around 700 cm$^{-1}$), that same bending motion at a higher excitation (weak feature around 1400 cm$^{-1}$) and its stretching motion (medium feature around 3300 cm$^{-1}$), forms the three main features of the HCN IR spectrum.

\begin{figure}[ht!]
\centering
\includegraphics[width=0.5\textwidth]{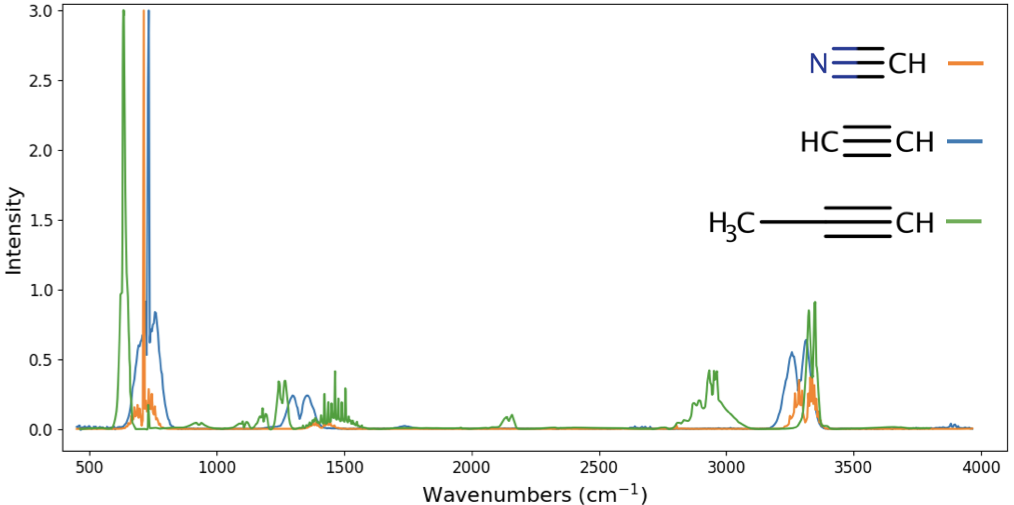}
\caption{\footnotesize{Comparison of the hydrogen cyanide (orange), acetylene (blue) and propyne (green) spectra.  Spectral data for hydrogen cyanide (HCN) is obtained from ExoMol\cite{tennyson2016exomol}, and the 
cross-sections for acetylene (C$_2$H$_2$) and propyne (C$_3$H$_4$) from NIST\cite{nist}. The three strongest features for HCN, C$_2$H$_2$ and C$_3$H$_4$ are associated with the same functional group, $\equiv$C-H. In the case of C$_3$H$_4$ (green), additional, weaker, features are visible, associated with the molecule's other functional groups. The spectral similarities between these three molecules can make it difficult to distinguish their contributions to an atmospheric spectrum.\label{deg}}}

\end{figure}

A single functional group, $\equiv$C-H, is responsible for all the main features of the HCN IR spectrum. As such, by seeing these features on a spectrum one may be detecting HCN; however, every other molecule with this functional group will also have these features, even if slightly modified. Consequently, at low resolution or with a narrow wavelength range, the detection of HCN will be vulnerable to degeneracies. For example, acetylene (C$_2$H$_2$), and propyne (C$_3$H$_4$) also have the same functional group, $\equiv$C-H, as HCN. There is no nitrogen in either of these molecules, but the $\equiv$C-H functional group does not depend strongly on what is on the other side of the carbon. The structural similarity between the three molecules leads to very similar spectra between them (see Figure \ref{deg}). 

The difficulty of distinguishing molecules with very similar spectra is not unique to the presented case
of acetylene, propyne and hydrogen cyanide. An RASCALL 1.0 estimation of the degeneracy of the $\equiv$C-H functional group suggests that there are approximately 806 molecules that share the functional group, and hence will show similar features between them. Examples include chloroethyne, fluoroethyne, propargyl alcohol, and cyanoacetylene. The unambiguous identification of molecules with the $\equiv$C-H functional group will therefore require either a much higher resolution, or the consideration of many more features beyond the strongest since they are all caused by the same functional group. At a minimum, identification of any of these molecules requires that some spectral data is available at all. Without RASCALL, this is currently true for only a handful of molecules with the $\equiv$C-H functional group.

It is important to note that a molecule's strongest feature is not necessarily that molecule's most distinguishing feature, and any molecular detection that does not take that into consideration will be vulnerable to false positives, false negatives and missasignments.
RASCALL 1.0 can not only simulate approximate spectra for any of the molecules that are targets for detection, but also highlight any degeneracies between possible detections. 
RASCALL 1.0 can then direct observations to regions of any given spectrum that can maximize the the potential for distinguishing molecules with similar functional groups.

\section{Discussion}\label{sec:discussion}
Modern molecular databases contain better spectra for more molecules than ever before. Between the libraries of historical spectral data (e.g., NIST\cite{nist}), and the continually updating modern databases of high-resolution molecular spectroscopy (e.g., HITRAN\cite{hitran2016}, ExoMol\cite{tennyson2016exomol}), we live in a golden age of reference spectroscopy. Together, these databases contain spectra and metadata for many hundreds of molecules. Nonetheless, only $\sim4\%$ of molecules from the 16,367 volatiles in the ASM list\cite{seager2016toward} have known spectral data, of any quality. For the remaining $96\%$ of molecules, there is no spectral information, and consequently it is currently impossible to remotely detect any of these volatiles.

The RASCALL project was created to address the paucity of molecular spectra in the literature. Given how difficult it is to obtain high-quality theoretical and experimental spectra, and taking into account the need for a vast spectral database of thousands of molecules, the current version of RASCALL provides a first approximation to the spectral features of all molecules in consideration. This initial version of RASCALL, RASCALL 1.0, can complement existing spectra for the few hundred molecules for which there is theoretical or experimental data, and can also serve as the sole source of spectral information for the thousands of molecules for which there are no data. The collection of spectral data provided by RASCALL contains 16,367 molecules - more than any other database; the RASCALL database will remain freely available, and will be continuously updated to improve coverage and accuracy.

The primary goal of the RASCALL project is to provide reliable spectra for all volatile molecules of interest to remote sensing on Earth and in the research field of astronomy. Given how ambitious this goal is, RASCALL will require many years of work before it reaches completion. RASCALL 1.0 takes into consideration the need for timely spectral data and primarily makes use of organic chemistry to create approximate spectra for thousands of molecules. 

The quality of the RASCALL 1.0 outputs depends on two major factors. First, the quality and completeness of the functional group dictionary. Second, the quality of the RASCALL 1.0 output depends on how well a molecule's spectrum can be approximated by the assessment of the molecule's functional group components. The successes and shortcomings of using solely functional group data to simulate spectra are discussed in this section, as well as the RASCALL 1.0 enhancements to the raw functional data.

 As the first of several RASCALL versions, RASCALL 1.0 produces spectra that are first order approximations with shortcomings in accuracy and completeness. These shortcomings are discussed in this section, in Appendix A, and will be addressed in future RASCALL updates (see section \ref{sec:future}).

\subsection{On the Completeness of the Functional Group Table}\label{sec:problems}

The quality of the RASCALL database is highly dependent on the completeness of the functional group dictionary, which contains information about the spectral properties of molecular vibrational modes. The RASCALL functional group dictionary is created from a functional group table which is routinely updated and maintained by the RASCALL team\cite{github}. For the creation of the RASCALL 1.0 database, most of the original functional group data is collected from tables in organic chemistry textbooks and online databases. These sources of functional group information are not exhaustive, nor can they be easily validated. Aggregating, assessing and enhancing a functional group table is therefore difficult and time-consuming, and there are many issues to consider. These issues are summarized below, and further details are provided in Appendix A.

First, aggregating sources. We collect functional group information from multiple sources and, where possible, combine them into averaged values. Even trustworthy sources have issues with under or over-generalization of a particular functional group's spectral behaviour; problems associated with this inevitable oversimplification can be partially mitigated by averaging between reliable sources.

Second, assessing data. Some sources have misassigned functional groups, or provide incorrect properties (e.g., intensities, positions). Where possible, functional group data have been verified by comparing to experimental spectra of molecules containing that functional group. Not all functional groups exist in a sufficiently large sample of molecules for this validation approach to work. In some cases, it was possible to assess functional group information as incorrect, and in such cases the affected entries were corrected and original data removed (see Appendix A for more details). 

Third, enhancing functional group data. In many cases, functional groups described in the literature were believed to be over-specific (e.g., functional group containing a fluoride atom instead of a general halogen), or over-generalized (e.g., a double bonded carbon, which can exist in dozens of functional group configurations). In these cases, new associated groups were tentatively created to represent more appropriately generalized, or specific, functional groups with similar properties. These are represented in the RASCALL spectral output by dashed lines (see, e.g., Figure \ref{hydro}). Additionally, functional groups will behave differently depending on the overall structural context of the molecule. This will mostly be addressed in RASCALL 2.0 but, for some molecules, RASCALL 1.0 already has some enhancement protocols in place (e.g., the muting of the C-H functional group by the $\equiv$C-H functional group).

For the RASCALL 1.0 functional group table, hundreds of functional groups, each with multiple spectrally active symmetries, have been catalogued. However, we would like to note that this functional group table is not exhaustive, and there are many entries that have not been included. In the RASCALL 1.0 functional group table, functional groups may be absent for the following reasons:
 \vspace{-2mm}
\begin{itemize}
\item{Missing functional groups: These functional groups correspond to active spectral features in a molecule's spectrum, but there have been no identifying measurements, or none have yet been found by the RASCALL team. The RASCALL 1.0 functional group table is the result of an extensive, though not exhaustive, literature review. As the RASCALL team extracts new functional group data from theoretical predictions and further literature searches, these will be added to the publicly available functional group table\cite{github}}

\item{Muted functional groups: In some cases, particular symmetries of a functional group, or the functional group as a whole, do not show significant, or measurable, spectral features. This spectral muting occurs because of interactions between the functional group and the other atoms and structures of the molecule. In some cases, the same functional group may be spectrally active in a group of molecules, but completely muted in another, depending on whether an additional functional group with masking properties is present. Where the cause of the muting is known, the functional group entry is annotated accordingly in the RASCALL code, and the RASCALL database updated.}

\item{Inactive functional groups: Theoretical estimates of a given molecule's total number of vibrational modes provide an upper bound to the number of functional groups a molecule can contain. However, many vibrational modes responsible for a given functional group symmetry do not result in a IR absorption (e.g., some are exclusively Raman active).}
\end{itemize}

The functional group table will remain a living document through the duration of the RASCALL project. As new functional groups are identified, and their properties become better defined, the functional group table will improve. The RASCALL 1.0 version of the functional group table, which, to our knowledge, is the most complete, publicly available database of functional group data, can be found online in the RASCALL 1.0 catalogue\cite{rascall_database}. Updated versions will be available in the RASCALL github repository\cite{github} throughout the duration of the RASCALL project. We are in the process of developing of a separate custom python computer code utilizing the modified rdkit substructure matching module\cite{landrum2006rdkit} that will identify all possible functional groups in any molecule. The code is based on a modified algorithm derived from that described in \cite{seager2016toward} and \cite{bains2012combinatorial}. The associated exhaustive list of all functional groups will be published alongside RASCALL 2.0.

\subsection{Bringing Approximate Spectra Closer to Reality}\label{enhancements}

The RASCALL 1.0 approach to simulating molecular spectra is intended as a first approximation. Each spectral feature predicted by functional group theory can be affected by neighbouring atoms and bonds, and influence from neighboring features. Consequently, variations between the behaviour of a functional group will lead to variations in the degree of agreement between approximate spectra created from functional group data and real spectra.

A wavelength region that is particularly difficult to simulate with RASCALL 1.0 is the spectral area between 500 and 1450 cm$^{-1}$, which is called the {\it{fingerprint region}}. Although RASCALL contains data on many functional group modes within this frequency range, the fingerprint region is often where a lot of deformation modes and combination bands are spectrally active, all of which can interact and blend with one another. Functional group theory is therefore weakest in the fingerprint region, a problem that is enhanced by the decreased reliability of organic chemistry data in this region.

The preliminary "skeleton" spectra provided by RASCALL 1.0 can be improved by increasing the accuracy of a molecule's band positions, shapes and intensities. RASCALL deals with these concerns individually, as follows:

{\bf{Band positions:}} The functional group table provides a range of frequencies where each symmetry of a functional group is expected to be spectrally active. RASCALL 1.0 takes the band centers for the symmetries to be the average of the upper and lower ranges in the functional group table. Where the RASCALL team finds the position of a particular symmetry to be consistently over, or under, estimated in the literature, the table is updated.  Future versions of RASCALL will further enhance band positions by creating sub-groups to account for variations within functional groups. Specifically, where the variation in positions of functional group features is clearly due to sub-classes of a functional group (e.g. functional groups containing different elements in the same periodic table column), new functional groups are created with higher structural specificity.

{\bf{Band shapes:}} For a handful of symmetries, an approximate band shape is reported in the literature (e.g. sharp, broad). This information is provided in the functional group table, but it is vague and often uncorroborated by additional sources. Consequently, band shape information from the literature is ignored by RASCALL 1.0. However, most programs that require molecular spectra, such as those modelling atmospheres, require molecular data in the form of cross-sections. To accommodate this input format, RASCALL 1.0 can apply a gaussian line function to add shape to the skeleton spectra, though this should be used carefully, as it is an unrealistic oversimplification of real molecular spectra. RASCALL 2.0 will provide preliminary band shapes using reliable quantum chemistry approximations, validated by experimental data where it is available.

{\bf{Band Intensities:}} Intensity information in the literature is particularly unreliable, both because of variations between sources, and also because of their vague qualitative descriptions (e.g., "strong",  "medium-weak"). To allow for a graphical depiction of the functional groups RASCALL 1.0 translated these qualitative intensities into values ranging from 1 to 3. Absolute intensities will be provided in future RASCALL versions but even the relative intensities provided by RASCALL 1.0 are assessed and modified where possible. As with band positions, where the RASCALL team finds the intensity of a particular symmetry to be consistently over, or under, estimated in the literature, the functional group table is updated. Finally, the majority of RASCALL 1.0 spectral data is either provided without temperature, or at room temperature, which provides only a rough approximation of the strength of spectral features for the molecules in the RASCALL database. Future versions of RASCALL will be able to integrate quantitative intensities, and these will be provided with temperature information.

\subsection{Future Versions}\label{sec:future}

RASCALL 1.0 exploits the consistency in functional group chemical properties to predict spectral features for thousands of molecules. The subsequent versions of RASCALL will aim to go beyond providing approximate band centers and qualitative intensities by improving the representativeness and breadth of molecular spectra in the RASCALL database. This will be achieved by combining tools and approximations from structural organic chemistry and quantum chemistry.

\subsubsection{Future RASCALL 1 Updates.}\label{future1}

The RASCALL 1 versions beyond 1.0 will maximize spectral coverage for as many molecules as possible by extracting all possible information from the available functional group data in the literature and experimental spectra. RASCALL 1 updates are summarized as follows:

1) An exhaustive list of functional groups will be obtained through a combination of computational chemoinformatics approaches and an in-depth search of the original literature and laboratory measurements of selected molecules containing novel functional groups. The subsequent "complete" functional group table can nonetheless be reassessed whenever new spectral data from reliable theoretical or experimental sources becomes available. 

2) The duplication of functional groups within a molecule will become visible in its spectrum. At present any repeated functional group (e.g., if a molecule has multiple methyl components) is spectrally indistinguishable from a single incidence of that functional group (see, e.g., Figures \ref{mol_example} and \ref{mol2_example}). We call these duplicated structures {\it{degenerate functional groups}}. In reality, a repeated functional group tends to express itself in a molecule's spectra by a shorter and broader feature with a double peak. These degenerate functional groups will be simulated in future RASCALL 1 versions.

3) Future RASCALL 1 updates will enhance approximate of molecular spectra from RASCALL 1.0 by predicting the systematic muting of functional group modes.  The most drastic intensity modification in RASCALL 1.0 was to address the selective muting of the C-H functional group (in molecules with a terminal $\equiv$C-H group, such as HCN and H$_2$C$_2$, where the C-H group is muted). New muting rules can be extracted through statistical analysis of muted functional group mode occurrences in measured and calculated spectra. Where consistent spectral muting patterns are identified, RASCALL 1 updates will implement new plotting conditions to reflect these patterns. It is worth noting that not all muted functional groups modes will necessarily be addressed as the process requires access to a large number of reliable experimental or theoretical spectra for multiple molecules with muted behavior.

\subsubsection{RASCALL 2.}\label{future2}

RASCALL 2 will make three significant improvements to the RASCALL 1 spectra:

1) RASCALL 2 will drastically change the way functional group data is processed into the program, so that specific behavior of functional group modes can be adapted to the individual molecule that contains that functional group. The current RASCALL 1.0. output only provides a zeroth order prediction of the spectral features of a functional group, but a functional group can have a variety of spectral expressions depending on which atom it is bonded to. Each functional group will be affected by its neighboring atoms and bonds, and the interference from neighboring features. The final spectral features for that functional group will consequently vary in their exact location, width, line shape and intensity. By estimating these effects and expanding the functional group analysis to non-CHNOPS elements, RASCALL 2 will be able to simulate full approximate spectra for an even larger number of molecules than the current RASCALL 1.0 database.

2) RASCALL 2 will provide band shapes for spectral features. To expand on the zero-width lines that RASCALL 1.0 currently outputs, RASCALL 2 will use a standard rigid rotor approximation to estimate the band surrounding the predicted band centers. Lines calculated using the rigid-rotor approximation will then be overlaid onto the band centers, and the band shape estimated using a functional form. The spacing and intensity of the lines can be predicted for each molecule by calculating its rotational constants. These require the total inertia of a molecule from the reference frame of the principal axis, which can be obtained by diagonalizing the Cartesian inertia tensor and retrieving its diagonal elements.

3) Intensities will initially be simulated by using the group contribution method (GCM) that scales band opacity with the number of repeated functional groups found in a molecule \cite{lopez2018monitoring}. Ultimately, the current qualitative intensities will be replaced by absolute values extracted from measured spectral features where they can be reliably estimated, and from theoretical fundamental band predictions where they can be simulated quickly (e.g., through DFT \cite{gross2013density} calculations such as B3LYP \cite{lee1988development}). At this stage, the RASCALL catalogue will contain temperature-dependent spectral data that can be transformed into approximate cross-sections and molecular opacities. We note that, at high temperatures, many RASCALL molecules will dissociate, but RASCALL 2 will nonetheless predict their spectra; we recommend that users only produce RASCALL spectra for temperatures in which a molecule is known to be stable.

Improving the band position, intensity and overall cross-sectional band shape will substantially improve the resulting spectrum for any molecule, and allow RASCALL 2 spectra to be used widely as an alternative to conventional molecular databases.

\subsubsection{RASCALL 3.}

RASCALL 3 will be the final version of the project, and it will include four main improvements from the previous versions.

{\bf{Uncertainties and resolution limits:}} Uncertainty estimates through identification of peaks in experimental measurements, and subsequent analysis of required resolution to distinguish between closely absorbing features of any given set of molecules.

{\bf{Inclusion of liquid and solid state data:}} The current functional group table contains data collected from infrared measurements of gas phase spectra. Many molecules have significant spectral features in their liquid and solid state, and individual features can correlate with known atom and bond substructures within the molecule. There is a rich collection of measurements of molecules in the solid phase that can be incorporated into RASCALL, such as studies of crystallized compounds with contain data on the spectral behavior of solid-state functional groups (e.g., \cite{sousa2013crystal}). This expansion beyond the gas phase will allow for RASCALL to be used, for example, in the interpretation of ices in Solar System bodies (e.g., \cite{cook2018composition}).  RASCALL 3 will collect all available functional group data for liquid and solid phase spectra, and add further data from theoretical extrapolations of known gas phase behavior.

{\bf{Inclusion of microwave, visible and ultra-violet data:}} The functional group data in RASCALL 1.0 is recorded in between 200 and 3700 wavenumbers, with the quality of the spectral coverage decreasing at the edges of this range. However, molecules with significant absorptions in the microwave (i.e. those with permanent electric dipole moments not containing the main rotation axis of the molecule) can have low frequency features associated with specific functional groups (e.g., internal rotations of the methyl group\cite{nair2006millimeterwave}). Likewise, molecules with rovibronic excitations can show significant features in the visible and ultra-violet that correspond to electronic excitations coupled with the allowed ro-vibrational modes of molecular functional groups. RASCALL 3 will expand its wavelength coverage by, when applicable, collecting and calculating all possible information of spectral activity beyond the infrared.

{\bf{Theoretical spectra:}} Using existing software, such as MOLPRO \cite{molpro} or Avogadro \cite{hanwell2012avogadro}, the geometry and electronic distribution of a molecule can be described. Then the position of the band centers can be calculated with a variety of accuracies, from reasonable (if speed is required) to excellent (if higher resolution is needed). By obtaining energy levels and associated transition intensities, RASCALL will be able to simulate spectra for a wide variety of pressure and temperature conditions. These simulations will require a significant level of understanding of the best combination of quantum chemistry theory and basis set, but are possible for a subsection of molecules, particularly if only approximate spectra is required. Quantum chemistry theories, such as VSCF \cite{bowman1986self}, DFT \cite{gross2013density} and MCTDH \cite{meyer1990multi} can already provide approximate spectra which can complement and, in some cases, supersede, the existing RASCALL predictions. However, such approaches are resource intensive and will only be possible for a small fraction of the molecules under consideration.


\section{Conclusion}
Given how experimentally challenging and computationally expensive it is to obtain accurate molecular spectra, there is great value in investing in intermediate approximate spectra that can be built upon as the accuracy requirements of astronomical studies increase. RASCALL 1.0 can already simulate the main spectral features of thousands of molecules, with applications in a wide range of fields. Future versions of RASCALL will combine state of the art quantum chemistry and functional group theory, providing increasingly accurate molecular spectra. 
The key to the remote detection of gases is molecular spectra. RASCALL aims to not just provide that key, but to do it at unprecedented computational speeds.

\vspace{10mm}
The authors thank Christopher Shea for his invaluable contributions to the RASCALL code, and Sarah Rugheimer for her contribution to the creation of the RASCALL acronym. We would also like to thank Fionnuala Cavanagh, Antonio Silva, Antonio P. Silva, Aaron Small, Megan Trivedi, Jenn Burt, Tom Whatmore, Martin Owens, Zhuchang Zhan, and Sukrit Ranjan for their useful support and discussions.

\newpage


\section*{Appendix A: Functional Group Table}\label{functional_appendix}

The raw functional group table is provided with the RASCALL 1.0 catalogue\cite{rascall_database} and in the RASCALL github repository\cite{github} with the file name "functionals\_dictionary.csv". The comma separated file can be opened as a spreadsheet, which is how we aggregate the functional group data obtained throughout the RASCALL project.

Whenever a mention of a functional group is found in the literature or through the analysis of theoretical and experimental spectra, it is translated into a SMILES\cite{weininger1988smiles} code that uniquely identifies the functional group class (e.g. C-H), symmetry (e.g. stretching) and its family (e.g. aldehyde). A report of a aldehyde C-H stretching vibration would therefore be recorded in the functional group table as "[H]C($=$O)[!\#1], stretch", where [!\#1] refers to any non-hydrogen atom.

Table \ref{tab:a} shows an extract of the functional group table "functionals\_dictionary.csv", with columns for the unique functional group code (FG SMILES), vibrational mode (FG Sym), frequency range (Freq$_{low}$, Freq$_{high}$), qualitative and quantitative band intensities (Int$_{Qual}$, Int$_{Quant}$), band shape (e.g., broad, sharp), source (literature, experimental data or RASCALL), functional group type (principal, general or associate) and functional group class (e.g., associated with the C$=$O bond), and molecular family (e.g., Alcohols, Aldehydes). We note that any functional group data that originates from the precursor to RASCALL 1.0\cite{atmos} has ATMOS as a source label.
	
\begin{table*}[]
\vspace*{-6mm}
\caption{\footnotesize{Sample of raw functional group table. Original sources \cite{irPDF, 05irPDF, ISAT}.}}\label{tab:a}
\centering
{\tiny
\begin{tabular}{clccccccccc}  
\hline
\hline
\noalign{\smallskip}
FG SMILES & FG Sym & Freq$_{low}$ & Freq$_{high}$ & Int$_{Qual}$ &Int$_{Quant}$ & Shape & Source & Type & Class & Family\\
\noalign{\smallskip}

\hline

\noalign{\smallskip}
$[$H$]$C($=$O)$[$!\#1$]$ & bend &  1380 & 1390 & m & UNK & UNK  & ISAT\cite{ISAT} &  principal & C-H & Aldehydes\\
$[$H$]$C($=$O)$[$!\#1$]$ & stretch &  1720 & 1740 & s & UNK & UNK  & Mult &  principal & C-O & Aldehydes\\
$[$H$]$C($=$O)$[$!\#1$]$ & stretch2 &  2690 & 2840 & m & UNK & UNK  & ir.pdf\cite{irPDF} &  principal & C-H & Aldehydes\\
$[$H$]$C($=$O)$[$!\#1$]$ & stretch2 &  2695 & 2830 & m & UNK & UNK  & ISAT\cite{ISAT} &  principal & C-H & Aldehydes\\
$[$H$]$OC &	bend &	650	& 770	& w	& UNK	& UNK &	ir.pdf	\cite{irPDF} & general &	O-H	& Alcohols/Phenols\\
$[$H$]$OC	& bend2	&1330&	1420&	w	&UNK&	UNK&	ISAT\cite{ISAT}&	general	&C-O	&Alcohols\\
$[$H$]$OC &	bend2&	1330	&1430	&w	&UNK&	UNK&	ir.pdf\cite{irPDF}&	general&	C-O	&Alcohols/Phenols\\
$[$H$]$OC	&stretch&	970&	1250	&w&	UNK	&UNK&	ir.pdf\cite{irPDF}	&general	&C-O	&Alcohols/Phenols\\
$[$H$]$OC	&stretch2&	1000&	1300	&w&	UNK	&UNK&	Mult &general	&C-O	&Alcohols/Esters\\
CC(C)$=$O&	stretch	&1710	&1720	&s	&UNK	&UNK	&05IRchart\cite{05irPDF}	&general	&C$=$O	&Ketones\\
CC(C)$=$O	&stretch	&1710	&1720	&s&	UNK	&UNK	&ir.pdf\cite{irPDF}	&general	&C$=$O&	Ketones\\
CC(Br)$=$O	&UNK	&UNK	&UNK	&UNK	&UNK	&UNK	&RASCALL	&associate & CC(Cl)=O	&UNK\\
P$[$H$]$	&bend&	950&	1250&	w	&UNK&	UNK&	ir.pdf	\cite{irPDF}&principal&	P-H	&phosphine\\
P$[$H$]$	&stretch	&2280	&2440	&m	&UNK	&sharp	&ir.pdf\cite{irPDF}	&principal&	P-H	&phosphine\\\hline
\vspace*{-3mm}
\end{tabular}
}
\vspace{4mm}
\caption{\footnotesize{Sample of molecular dictionary, in the input format for RASCALL 1.0. The symbol [!\#1] represents any non-hydrogen atom.}}\label{tab:b}
\centering
{\tiny
\begin{tabular}{cclclclcl}  
\hline
\hline
\noalign{\smallskip}
Mol SMILES & FG$_1$ & n(FG$_1$) & FG$_2$ & n(FG$_2$)  & FG$_3$ & n(FG$_3$) & FG$_4$ & n(FG$_4$)  \\
\noalign{\smallskip}

\hline

\noalign{\smallskip}

'C(C1)(C1F)(CC)'&	'C[H]'	&9 &	'[H]C([H])([!\#1])[!\#1]'	&2 &'CF'	&1 &'[H]C([H])([H])[!\#1]'	&1 \\
'IC(I)[AsH]CC' &	'C[H]'	&6 &	'[H]C([H])([!\#1])[!\#1]'	&1 &'CI'	&2 &'[H]C([H])([H])[!\#1]'	&1 \\
'O=PCCl' &	'C[H]'	&2 &	'[H]C([H])([!\#1])[!\#1]'	&1 &'CCl' & 1 & 'P$=$O' & 1\\
'[AsH2]C(Br)=C(Cl)Cl' &	'C$=$C' & 1 &	'[!\#1]C([!\#1])$=$C([!\#1])[!\#1]' &1 &'CCl' & 2 & 'CBr' &1\\
'CCCCCBr' &	'C[H]'	&11 &	'[H]C([H])([!\#1])[!\#1]' & 4 &'CBr' & 1 &'[H]C([H])([H])[!\#1]' & 1 \\\hline
\vspace*{-3mm}
\end{tabular}
}
\vspace{-2mm}
\end{table*}

Where possible, multiple sources for the same functional group mode are obtained and inspected for coherence of spectral data ({\it{Mult}} in the Source column of the functional group table). Coherent functional group data with multiple sources is considered the most reliable. The RASCALL code averages the values of the intensities and the frequencies provided by multiple sources for the same functional group mode. Where it is not possible to confirm that similar functional group data from different sources are referring the the same functional group, new temporary functional groups are created to include all the data. If, upon inspection of the RASCALL simulated spectra, the similar functional group data are found to belong to the same spectral feature, then the multiple functional groups are collapsed into a single entry.

 An additional table, called "functionals\_dictionary\_update.csv", also found in the GitHub repository\cite{github}, contains a more complete selection of functional group data but with a portion of contents still under revision. We recommend that users primarily refer to the table in "functionals\_dictionary.csv", but the "functionals\_dictionary\_update.csv" document contains all of the functional group information as it is directly processed from the literature. 
 
 Common issues with the raw data in the functional group tables include:
 
 {\bf{Missassigned functional group labels:} }For example, the C$=$C asymmetric stretch as reported by Tarek Ghaddar at the American University of Beirut\cite{irPDF}. We plotted 118 molecules with the C$=$C functional group that have spectra from both RASCALL 1.0 and NIST. In every single one of them, the predicted spectral feature from the C$=$C asymmetric stretch, supposedly a strong absorption at 1950 wavenumbers, was not correctly predicted. This particular mode, the C$=$C asymmetric stretch, was consequently deleted from the RASCALL functional group table. It is possible that the feature should have been assigned instead to the C$=$C$=$C functional group, which does absorb, quite strongly, at 1950 wavenumbers.

{\bf{Inconsistent absorptions:}} For example, 96 molecules in NIST have the $[H]$OC functional group. Multiple sources report significant absorptions associated with multiple symmetries (i.e. modes) of the $[H]$OC functional group. However, almost all molecules that contain this functional group and have available experimental spectra show at most weak absorptions associated with the $[H]$OC functional group. We have temporarily reduced the intensity of all spectral absorptions associated with the $[H]$OC functional group until further analysis.
 
 {\bf{Mislabelled and missing symmetries:}} The labels in the RASCALL 1.0 functional group table labels support only bending and stretching (symmetric and asymmetric) modes. However, we know from inspection that the modes described in the functional group table include hot or combination bands, as well as torsional, rocking, scissoring, or wagging motions. These are found under the placeholder labels ending in integers (e.g., {\it{bend3}}). These labels should be considered temporary, and will be replaced with their true symmetry labels as the RASCALL team is able to confidently assign their identity.

In some cases, there is available literature for functional groups containing more than six non-hydrogen atoms. Given that the current RASCALL molecular database only contains whole molecules with $\leq$6 non-hydrogen atoms, these large functional groups were dismissed. Extending the ASM database beyond 6 non-hydrogen atoms is not very difficult and may be undertaken in the future (see \cite{seager2016toward} for more details). Any extension to the ASM list of molecules, or the addition of any other molecular list with SMILES codes, can be easily processed through the RASCALL code to extend the RASCALL database.

\section*{Appendix B: Molecular Dictionary}

The current molecular dictionary collects information on the identity, and number of instances, of functional groups in ASM molecules. The identification of functional groups and the enumeration of the instances of each of the functional groups in ASM molecules was carried out using a custom software built using the RDKit Python toolkit\cite{landrum2006rdkit}. The structures of ASM molecules and functional groups are represented in SMILES/SMARTS notation and are converted by the software to RDKit internal mol format before the structural matching procedure. The software searches for the structural match between a functional group structure and a structure within the target ASM molecule. The match has been found only if the structural match in ASM molecule is exactly as that of the query structure of the functional group. 

Table \ref{tab:b} shows an excerpt from the molecular dictionary used by RASCALL 1.0, where each molecule is described as a python dictionary key (represented with a SMILES code) with its value as a list of tuples containing the functional group code (also in SMILES notation) and incidence number of each identified functional group, n(FG$_i$).

The RASCALL 1.0 molecular dictionary is available as supplementary material to the RASCALL 1.0 database\cite{rascall_database}, in a Python serialization data format (pickle). Updated versions of the molecular dictionary can be found in the RASCALL GitHub repository\cite{github}. We note that every updated to the functional group table requires the molecular dictionary to be updated.

\section*{Appendix C: RASCALL Code}\label{code}

The RASCALL 1.0 code is written in python 3, with modular programs for processing the functional group data, the molecular dictionary, all plotting functions, wavelength/intensity filtering and all enhancements of functional group data. The central code accesses the necessary modules to readily list or plot any selected molecule, all molecules with a particular functional group, and any molecule of a specified family (e.g., hydrocarbons, halocarbons).

RASCALL 1.0 can list or plot the following: a single molecule with its associated functional groups and spectral properties; a list of molecules containing a particular functional group; all molecules belonging the hydrocarbon or halocarbon families; molecules with a specific atomic element; molecules with significant absorptions within a given frequency range, over a given intensity threshold; molecules that are absorbers in the windows of preset atmospheres (e.g. methane-rich, archean Earth). Some of these parameters can be listed or plotted through command line user arguments and some from within the RASCALL 1.0 modules. Future RASCALL updates will focus on improving the user interface for all features.

All RASCALL arguments require SMILES format\cite{weininger1988smiles}. To improve accessibility to the RASCALL code and database, a file has been provided on GitHub which provides alternative identifiers for each molecule in the RASCALL database. This file, called "RASCALL$\_$Molecule$\_$Identifiers" contains the SMILES code for each RASCALL molecule, alongside other common identifiers (e.g., IUPAC chemical names, canonical SMILES, molecular formula, InChi Code/Key).
 
All RASCALL 1.0 modules are available on GitHub\cite{github} and can be used to produce the most up-to-date RASCALL database at any point. Every iteration of the RASCALL code and its inputs (i.e. the molecular dictionary and the functional group table) will remain open source in perpetuity. However, as the RASCALL code will remain under development until its final version, users may require assistance to access specific features; if necessary, the RASCALL team may be able to provide installation and implementation support. In the GitHub repository we also provide a virtual python environment, under "pythonenv", so that users can create a temporary, isolated, environment with the exact python dependencies used by the RASCALL code.

\section*{Appendix D: RASCALL Molecular Catalogue}\label{database}

The preliminary RASCALL catalogue is freely accessible online and contains spectral information about 16,367 molecules\cite{rascall_database}. The catalogue contains a functional group dictionary (see Appendix A), the molecular dictionary (see Appendix B), the molecular database, and an associated README file. In the  molecular database there are folders with the name of each molecule in SMILES format \cite{weininger1988smiles}. In each molecule's folder there are two files, one containing the RASCALL skeleton spectra data, and one with the molecule's functional group information.

Please be advised that the RASCALL 1.0 catalogue of molecular spectra presented in the current repository \cite{rascall_database} is a preliminary database. The RASCALL code\cite{github} can always be used to output the most up-to-date RASCALL database locally, but new versions of the RASCALL database will only be made available online whenever major improvements in RASCALL occur. The GitHub repository also contains the document "RASCALL$\_$Molecule$\_$Identifiers", which has a list of commonly used identifiers for all RASCALL molecules.

It is recommended that users contact the first author for technical support and clarification before using the RASCALL data as molecular input for spectra simulations.




\end{document}